\documentclass[showpacs,amsmath,amssymb,a4]{revtex4}
\usepackage{graphicx}
\usepackage{dcolumn}
\usepackage{bm}
\newcommand{\Zeff}{Z_{\text{eff}\,}}
\newcommand{\kapeff}{\kappa_{\text{eff}\,}}

\newcommand{\Pocm}{P_{\text{ocm}}}
\newcommand{\Pcoll}{P_{\text{coll}}}
\newcommand{\Pmicro}{P_{\text{micro}}}
\newcommand{\be}{\begin{equation}}
\newcommand{\ee}{\end{equation}}
\newcommand{\Zbare}{Z_{\text{bare}}}

\begin{document}
\title{Testing the relevance of effective interaction potentials
between highly charged colloids in suspension}

\author{J. Dobnikar$^{1,2}$,
R. Casta\~{n}eda-Priego$^3$, H.H. von Gr\"unberg$^2$, E. Trizac$^{4,5}$}
\affiliation{(1) Jozef Stefan Institute, Jamova 39, 1000 Ljubljana, Slovenia \\
(2) Institut f\"ur Chemie, Karl-Franzens-Universit\"at,
Heinrichstrasse 28, 8010 Graz, Austria,\\ 
(3) Instituto de F\'isica,
Universidad de Guanajuato, 37150 Leon, Mexico,\\ (4)
CNRS; Universit\'e Paris-Sud, UMR 8626, LPTMS, Orsay Cedex, F-91405\\
(5) Center for Theoretical Biological Physics, UC San Diego,                    
       9500 Gilman Drive MC 0374 - La Jolla, CA  92093-0374, USA}

\date{\today}
\begin{abstract}
Combining cell and Jellium model mean-field approaches, Monte Carlo together with
integral equation techniques, and finally more demanding many-colloid
mean-field computations, we investigate the thermodynamic behavior,
pressure and compressibility of highly charged colloidal dispersions,
and at a more microscopic level, the force distribution acting on the
colloids.  The Kirkwood-Buff identity provides a useful probe to
challenge the self-consistency of an approximate effective screened Coulomb (Yukawa) 
potential between colloids. Two effective  parameter models are put to the test: 
cell against renormalized Jellium models.  
\end{abstract}
\maketitle

\section{Introduction}

Colloidal suspensions contain mesoscopically large particles from the
$nm$ to $\mu m$ size regime, the colloids, together with small solvent
and solute molecules. These molecules are orders of magnitude smaller than
the colloids, and still, they heavily influence the interactions between the
colloidal particles. Such microscopic species also significantly
affect the thermodynamics of the system. However, they often cannot be
directly visualized and most experimental techniques such as small
angle X-ray or neutron scattering, probe the colloidal degrees of
freedom only.  This stems from the wide separation of characteristic
time and length scales between microscopic and mesoscopic degrees of
freedom, which therefore suggests to develop {\em effective}
approaches for the colloidal particles, integrating over
their microscopic counterparts, that follow adiabatically.

The focus of this manuscript will be on the behaviour of
charge-stabilized colloidal suspensions
\cite{luc,Hansenrevue,Levin}. In the subsequent analysis, we will
further restrict ourselves to a mean field treatment where the
correlations between microions are discarded. The resulting
Poisson-Boltzmann (PB) description is valid in aqueous solvents under
usual conditions of temperature and for monovalent microions, because
existing colloids do not allow to reach the high Coulombic couplings
that are required to observe deviations from mean-field
\cite{Levin}. In the following, we therefore focus on the more special
case of charged colloids in a simple 1:1 electrolyte, adopt the PB
approach, and treat the micro-ions as point-like particles dissolved
in a continuum of solvent molecules.

Even within such a simplified framework, the solution corresponding to
$N_c$ interacting colloids is a difficult problem from a numerical
point of view \cite{Fushiki,Madden}.  In principle, one has to compute
the density distribution of microions in space
$\rho_{\text{micro}}({\bm r})$ for any given colloidal
configuration. First one has to solve the PB equation around a large
collection of $N_c$ colloids. One can then compute the corresponding
stress tensor to obtain the force felt by each colloid, from which the
colloidal configuration of the next time step can be found.  This
defines the loop that has to be iterated -- a numerically rather
demanding task \cite{Fushiki,Rzehak,JUREEPL}. It is therefore still of
interest and common practice to map the multi-component Hamiltonian of
the charge-stabilized colloidal suspension onto a one-component model
(OCM). In the OCM the colloidal particles interact via {\it effective}
pair-potentials $u_{\text{eff}}$ \cite{luc,Hansenrevue,Levin} -- {\it
effective} in the sense that these pair-potentials reflect not only
the true pair-potential between the ''naked'' colloids, but also the
indirect effect that all the micro-ions and solvent molecules have
onto the interacting colloids. As an advantage of the OCM, the number
of degrees of freedom is significantly reduced: If each colloid has
$Z$ counter-ions, then a system with $N_c$ colloidal particles has at
least $Z$x$N_c$ interacting species (not to speak of any additional salt
ions or the solvent molecules), which in the OCM are reduced to just
$N_c$ interacting colloids.

The effective pair-potentials between charged colloids are usually
Yukawa like (screened Coulomb, see Eq.~(\ref{eq7}) below), but with
effective charge and screening parameters whose dependency on the
colloid density makes the whole pair-potential
density-dependent. There is no rigorous way how to ``derive'' these
effective parameters from first principles. However, a large number of
rather sophisticated recipes to determine effective parameters can be
found in literature \cite{luc,Hansenrevue,Levin}; but no matter how
sophisticated a scheme is, an heuristic element within these theories
can never be avoided. In section~\ref{sec:2models} we will briefly
describe two rather simple approaches to compute effective
Yukawa-parameters, in the following referred to as the ''PB cell
model'' (see e.g. \cite{deserno}) and the ''Jellium model''
\cite{trizaclevin}.

Combining various numerical simulation techniques (Monte Carlo,
integral equation, and $N_c$-body mean-field computations), we here
investigate the performance of these effective Yukawa potentials and
the two effective-charge theories, the PB cell model and the Jellium
model.  We will perform these tests in three different ways: (i) we
first investigate if the form of the effective Yukawa potential is
able to reproduce the correct distribution of forces felt by the
colloids in various configurations. Such a comparison, that will be
addressed in section \ref{sec:manybody}, requires the solution of PB
problem around $N_c$ colloids. One may anticipate that in situations
of high salt content, the Yukawa approximation will be operational,
while the opposite low salt regime deserves more attention. (ii) We
then study the compressibility $\chi$ of the suspension as a function
of the colloid density. Both the PB-cell and the Jellium model do not
only provide us with effective parameters for the Yukawa
pair-potential, but directly predict the pressure and compressibility
of the suspension. The Kirkwood-Buff relation now allows to check the
thermodynamic consistency of the effective-charge models; it relates
$\chi$ to $S(0)$, the infinite wavelength limit of the colloid-colloid
structure factor. While $\chi$ follows directly from the two effective
charge models, $S(0)$ is related to the structure of the suspension
which can be obtained from the colloid-colloid pair-correlations
calculated using the effective Yukawa forces of the OCM. As these
Yukawa forces again require the effective parameters of the two
effective charge models, the Kirkwood Buff relation can be used for a
stringent test of the consistency of the effective charge models under
scrutiny.  Such a consistency check is performed in
section~\ref{sec:structthermo}.  (iii) The most direct way to check
the quality of an approximation is to compare it to the results of
more rigorous approaches. In the third main section of this paper
(section~\ref{sec:primitive}), we compute osmotic pressure,
compressibility and pair-correlations of the suspension for the two
effective-charge models and compare the results to primitive model
calculations.

%%%%%%%%%%%%%
\section{Two models for effective charges and screening lengths}
\label{sec:2models}

We start by outlining the two effective charge models that we here are
concerned with. The idea behind using effective parameters to
incorporate effects of non-linear screening in a pair-potential based
otherwise on linear theory, is explained and reviewed in
Ref.~\cite{bell00}. 

Although the effective potential has an unambiguous definition, there
is no rigorous {\em operational} route to construct this object and it
is common belief, under scrutiny here, that a Yukawa form
\cite{luc,Hansenrevue,Levin}
\begin{equation}  
\label{eq7}
\beta u_{\text{eff}}(r) = Z_{\text{eff}}^{2} \lambda_B \left(
\frac{\exp(\kappa_{\text{eff}}\, a)}{1 + \kappa_{\text{eff}}\,
a} \right)^2 \frac{\exp(-\kappa_{\text{eff}}\, r)}{r}
\end{equation}  
provides a reasonable description. Here $a$ is the radius of the
colloid (assumed to be a sphere), $\lambda_B$ is the Bjerrum length,
$kT=\beta^{-1}$ is the thermal energy, while $Z_{\text{eff}}$ and
$\kappa^{-1}_{\text{eff}}$ are the effective charge and screening
length.  Such a ``DLVO''-like expression \cite{verwey} would
accurately reproduce the large distance interaction of two colloids
immersed in a salt sea \cite{luc,Hansenrevue,Levin} but it should be
kept in mind that it has to fail at short distances \cite{luc}.  One
should also keep in mind that even if one had the perfect effective
potential for two colloids, one can not be sure that it is also the
appropriate effective pair-potential for a concentrated suspension of
colloids, because in general these effective pair-potentials cannot be
superposed. Many-body interactions between the colloids must be taken
into account \cite{brunner} as they can have a considerable impact on
the structure of the suspension \cite{russ05}. We also remark, that in
the salt free case, the validity of (\ref{eq7}) is not obvious.  Other
limitations of the concept of effective pair-potentials are discussed
elsewhere \cite{luc,Hansenrevue,Levin,dijkstra}.

In appendix \ref{app:effchrg} we recapitulate the essentials of the
two empiric models (Jellium and PB-cell model) used here to determine
$Z_{\text{eff}}$ and $\kappa^{-1}_{\text{eff}}$ in the Yukawa
pair-potential. As emphasized in the introduction, effective
parameters can also be used to predict the osmotic pressure of the
suspension.  We define the osmotic pressure from the pressure $P$ and
the reservoir pressure $P_{\text{reservoir}}$, as \be \Pi
\,=\,P-P_{\text{reservoir}} = P-2 c_s kT\:,
\label{eq:Pi}
\ee 
where $c_s$
denotes the (monovalent) salt concentration in the
reservoir. Furthermore, the osmotic isothermal compressibility $\chi$
is defined via Eq.~(\ref{eq:Pi}) as
\begin{equation}  \label{eq4}
\chi^{-1}=\frac{\partial \beta \Pi}{\partial \rho_c}\biggl|_{T,\text{salt}} \:,
\end{equation}  
where the derivative with respect to the colloid density $\rho_c$ is
taken at constant salt chemical potential (i.e. constant $c_s$).
Applying these two definitions, the osmotic pressure of the suspension
can be related to the effective parameters (see
appendix~\ref{app:effchrg})
\begin{equation}  
\label{eq14}
4 \pi \lambda_B a^2 \beta \Pi_{\text{micro}} =
\kappa^{2}_{\text{eff}}\,a^2 - \kappa^2a^2 
\end{equation}  
with $\kappa^2=8 \pi \lambda_B c_s$. The suffix ``micro'' is explained
further below. The compressibility then reads
\begin{equation}  
\label{eq14a}
\chi^{-1}=\frac{a}{3 \lambda_B}
\frac{\partial \kappa^{2}_{\text{eff}}\,a^2}{\partial \eta}
\end{equation}
in which $\eta=4 \pi \rho_c a^3 /3$ is the colloid volume-fraction. In
these expressions $\kappa^{2}_{\text{eff}}\,a^2$ is given either by
the PB-cell expression, Eq.~(\ref{eq11a}) and ~(\ref{eq11b}), or the
corresponding expression for the Jellium model, Eq.~(\ref{eq9}).

While the Jellium model is implicitly based on the assumption that the
colloid-colloid pair distribution function $g(r)\simeq 1$, the PB cell
model takes the opposite view and assumes a $g(r)$ corresponding to a
crystalline colloidal configuration. This latter assumption rests on
the observation that the repulsively interacting colloids arrange
their positions such that each colloid has a region around it which is
void from other colloids and which looks rather similar for different
colloids. In other words, one model assumes a regular arrangement of
the colloids in the suspension and thus a rather solid-like structure,
while the other model assumes no structure at all -- a situation
typical of fluids at low density. In that sense, the two models chosen
are complementary to each other.

The effective charges as a function of the bare colloid charge
$\Zbare$ go to $\Zbare$ for low bare charges, and to a saturated
effective charge $Z_{\text{eff}}^{sat}$ for high charges, and can
therefore be roughly approximated by
\begin{equation}  \label{eq15}
Z_{\text{eff}}(\Zbare) \approx 
\left\{
\begin{array}{l@{\hspace{2cm}}l} \Zbare & \Zbare< Z_{\text{eff}}^{sat} \\
Z_{\text{eff}}^{sat} & \Zbare \ge Z_{\text{eff}}^{sat} \end{array} 
\right.
\end{equation}  
Except in section \ref{sec:primitive}, we will mostly concentrate on
the saturated regime that is most suited to describe colloidal
suspensions \cite{trizac1} (for a discussion of the crossover
behaviour between low and high bare charges see \cite{Aubouy} for
symmetric electrolytes and \cite{Tellez} for 1:2 and 2:1
electrolytes).  By choosing in all our calculations a $\Zbare$ large
enough to ensure saturation of the effective charges, we can thus get
rid of the additional parameter $\Zbare$.

We have plotted the effective saturated charges as a function of
$\eta$ in Fig.~(\ref{fig1}) of the appendix. One can identify a regime
in $\eta$ where salt-ions dominate the screening, leading to effective
charges that are practically independent of the volume-fraction. At
other values of $\eta$ however counter-ions outnumber the salt-ions;
this is the regime where screening is dominated by the
counter-ions. As the number of these counter-ions depends on the
number of colloidal particles, the effective charges become
$\eta$-dependent when the counter-ions dominate the screening.  One may
estimate from Eq. (\ref{eq9}) of the appendix the volume-fraction
$\eta^*$ where one may expect to find the crossover between both
screening regimes,
\begin{equation}
\label{eq:etastar}
\eta^* \simeq \frac{\kappa^2 a^3}{Z_{\text{eff}} \lambda_B}\:.
\end{equation}
We remark that this threshold value $\eta^*$ was derived within the
Jellium framework. We have no clear definition like this for the cell
model and we have determined the $\eta^*$ for the cell model
empirically (see section \ref{sec:KB}). Empirically, as will be
discussed in the section \ref{sec:KB}, the crossover volume fraction for 
the cell model is found to be much smaller, $\eta \approx 0.2 \eta^*$.

Working in the regime of saturated effective charges, we are left with
 in total only two independent input parameters: the volume fraction
 $\eta$ and the salt concentration $\kappa a$ of the salt reservoir.
 Multiplying $Z_{\text{eff}}^{sat}$ with $\lambda_B/a$ and the osmotic
 pressure with $4 \pi \lambda_B a^2$, as we have done in
 Eqs.~(\ref{eq9}),(\ref{eq12}),(\ref{eq14}), we avoid $\lambda_B/a$ as
 an additional independent parameter.  However, the colloid-colloid
 pair correlation function $g(r)$ does not exhibit the same scaling
 behaviour as $\Pi_{\text{micro}}$ so that it is important to precise
 which value of $\lambda_B/a$ has been used in the simulations. Unless
 otherwise specified, we have considered $\lambda_B/a = 0.01$, a
 reasonable value for colloidal systems in an aqueous environment
 (smaller values are also met in experiments).  As for $\eta$ we have
 stopped our calculations at high volume fraction when the liquid
 started to solidify, while for $\kappa a$ we have worked in between
 two extremes, $\kappa a =1.5$ (high-salt regime) and $\kappa a =0.0$
 (no salt).

\begin{figure}
%\mbox{}\\[1cm]
\includegraphics[width=0.7 \textwidth]{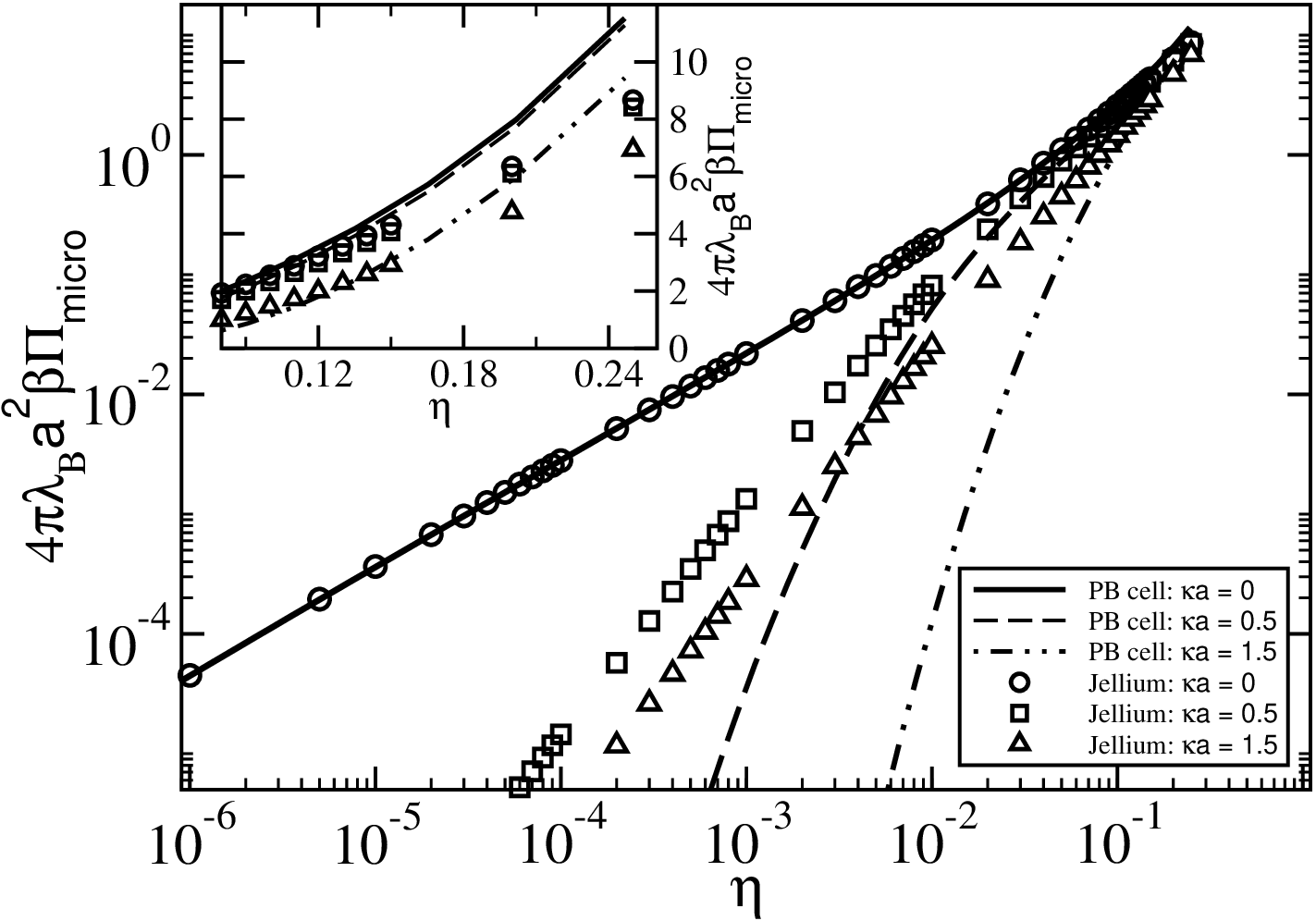}
\caption{\label{fig2} Volume-fraction dependence of the reduced
osmotic pressure $\Pi_{\text{micro}}$ according to the Jellium model
(symbols) and the PB-cell-model (lines). Symbols and lines as defined
in the legend. The scale in the plot is logarithmic, while the
inset shows the same on a linear scale.}
\end{figure}
Fig.~(\ref{fig2}) compares the pressure prediction
$\Pi_{\text{micro}}$ of both models. The curves are based on
Eq.~(\ref{eq14}) in combination with (\ref{eq11a}), (\ref{eq11b}), and
(\ref{eq9}). In the no-salt case both models have the same
low-dilution behavior, but different limiting behavior in the presence
of salt ions where we have an algebraic decay for the Jellium, but an
exponential one for the PB cell model.  The agreement between Jellium
and PB-cell is excellent in the no-salt case, up to volume fractions
around $0.1$ which is remarkable in view of the differences in the
effective charges.

Before proceeding we wish to emphasize that many more effective-charge
approaches can be found in literature, including DFT-based schemes
(e.g. \cite{Stevens}), Monte Carlo solutions of the cell model
(e.g. \cite{jonsson,groot}), and other mean-field models with various
criteria for the effective charge (see reviews in
\cite{bell00,dijkstra}).  We here have selected both the Jellium and
the PB-cell models because they are complementary to each other and
because they are probably the two most practical effective-charge
models; they are sufficiently accurate, well-established in
literature, quite simple and rather straightforward to implement (see
appendix of \cite{trizacLang}).

\section{Test I: Many-body forces versus Yukawa forces}
\label{sec:manybody}

We are now in the position to perform the first test. We recall that
the interactions among colloids in a suspension are mediated by the
microions and are therefore in origin complex many-body
interactions. Integrating the microions out, the force on any given
colloid depends on the positions of all other colloids in the
system. It is not straightforward that this force can be written as a
sum of pairs only. In principle, one has to sum over pairs, triple,
... and over all many-body configurations. To test how important the
higher many-body contributions are, the true many-body forces have
to be compared to the forces obtained by summing up the effective pair
interactions \cite{LowenMulti}. This is done in this section.

\begin{figure}
\mbox{}\\[1cm]
\includegraphics[width=0.485\textwidth]{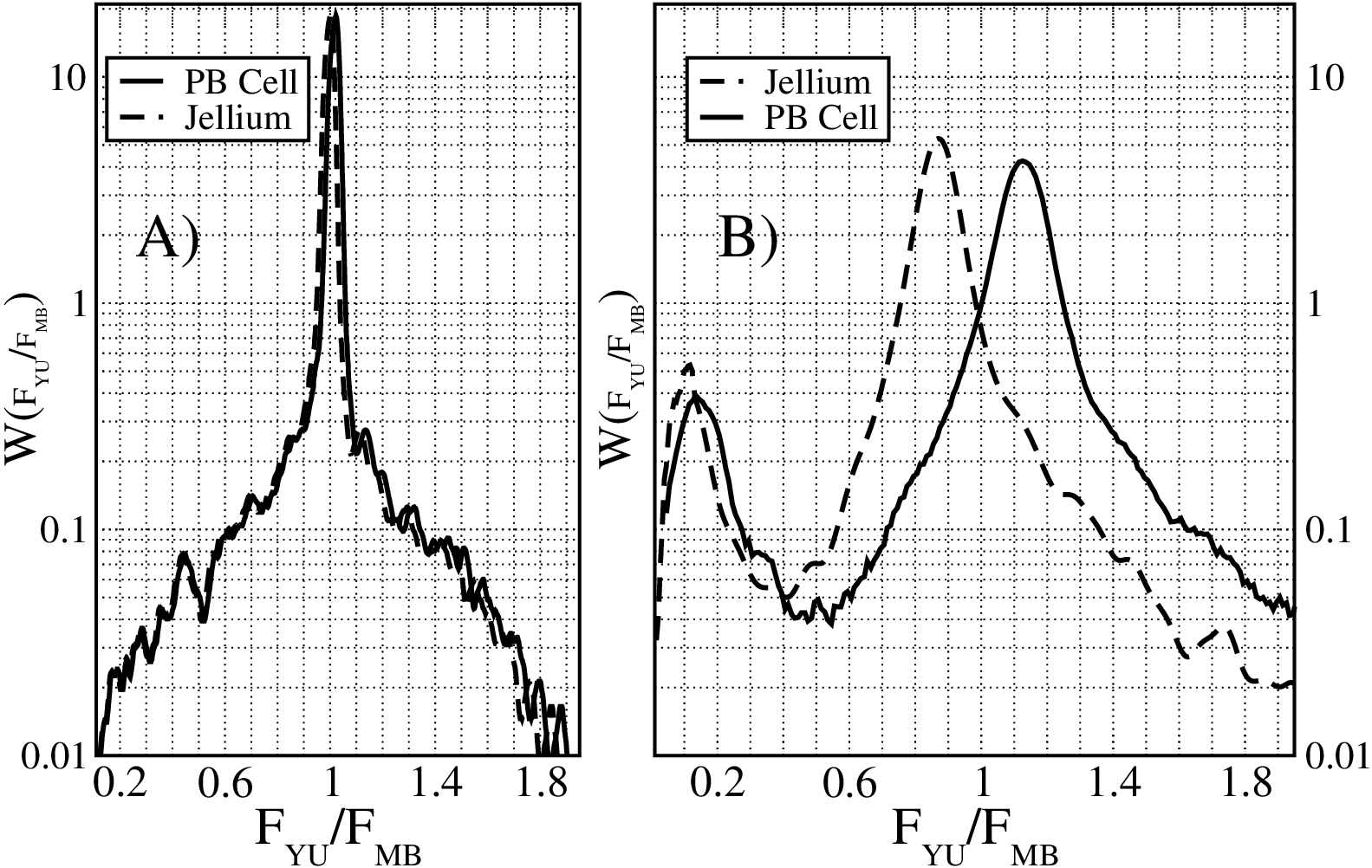}
\hfill\includegraphics[width=0.45 \textwidth]{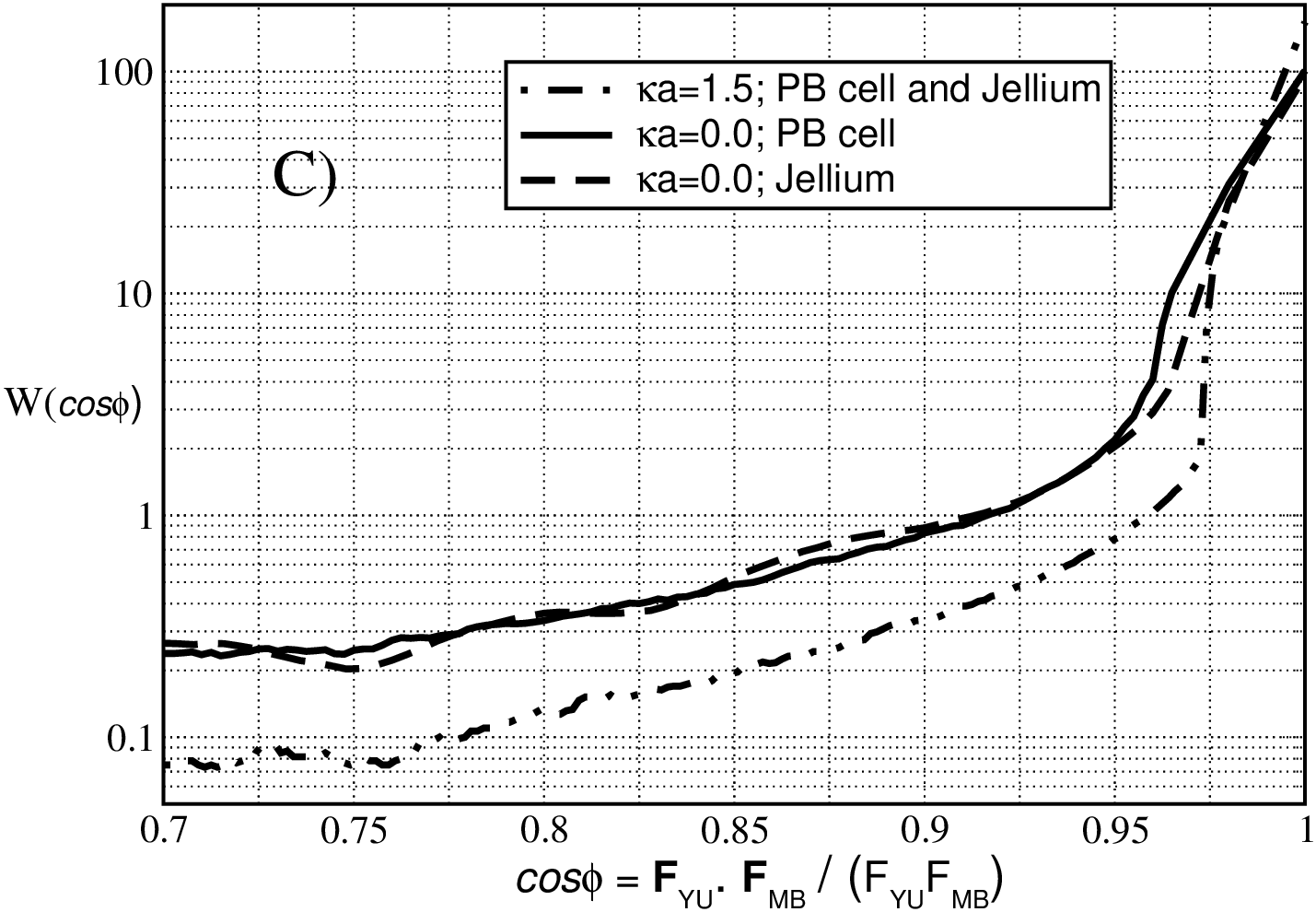}
%\mbox{}\\[2.5cm]
\caption{\label{fig7} Histogram of ratio between the moduli of Yukawa
pair forces $F_{YU}$ and the correct many-body forces $F_{MB}$ for
{\bf A)} high salt concentration ($\kappa a =1.5$) and for {\bf B)} no
added salt. A few particle configurations are obtained from MC
simulations with 4000 and 6000 particles at the volume fraction
$\eta=0.0001$.  $F_{YU}$ is evaluated with both effective parameters
approximations, the PB cell and the Jellium. $F_{MB}$ is obtained by
solving the nonlinear Poisson-Boltzmann equation in the given
colloidal configurations. The bare charge is such that  
$Z\lambda_B/a =50$, which
is large enough to effectively reach the saturation regime. We have
also performed the calculations at larger volume fraction $\eta=0.01$
and found very similar distributions as the ones shown here at
$\eta=0.0001$. 
{\bf C)} The distribution of relative angles between
the correct solution ${\bm F}_{MB}$ and the two effective pair
pictures. All graphs are presented in a linear-log scale.}
\end{figure}

We arbitrarily selected a few typical colloidal configurations from a
MC simulation of 4000 Yukawa particles and solved the non-linear
Poisson-Boltzmann equation in the region between the colloidal spheres
of this colloidal configuration using our multi-centered
Poisson-Boltzmann solver described in detail in \cite{Rzehak}. As in
\cite{Rzehak} we integrated the stress-tensor around each colloidal
sphere to obtain the force acting on each colloid. These forces ${\bm
F}_{MB}$ -- which include all many-body forces acting on the
individual colloid -- can now be compared to the forces ${\bm F}_{YU}$
one obtains by summing pair-forces derived from Yukawa-potentials,
Eq.~(\ref{eq7}), with effective PB-cell or Jellium parameters. We have
evaluated the ratio of the magnitudes $F_{YU}/F_{MB}$ and the (cosine)
angles between the forces ${\bm F}_{YU}\cdot{\bm F}_{MB}/F_{YU}F_{MB}$
for each particle and have plotted the corresponding histograms, that
have been averaged over a few configurations. The results of these
calculations are shown in Fig.~(\ref{fig7}) for the salt-free case and
for the high salt case, $\kappa a = 1.5$.

We see from Fig.~(\ref{fig7}.A) that the ratio between the forces is
very close to 1 and has a narrow delta-like distribution in the high
salt case, as we should expect, since the Yukawa one-component picture
has already proven to be applicable at such parameters.
Interestingly, in the no salt case, the force ratio shows a broader
distribution but is still quite peaked, although not centered at
unity, see Fig.~(\ref{fig7}.B). The forces based on the Yukawa
one-component picture are therefore on average larger (cell) or
smaller (Jellium) than the correct many-body forces.  We may then
anticipate that when used in the OCM, Jellium parameters will
underestimate the structure, while cell data will lead to an
overestimation (see Fig. \ref{fig6}B).  Despite these deficiencies,
both models seem to reproduce the forces reasonably well, which is
especially true for the angular distribution displayed in
Fig.~(\ref{fig7}.C). An interesting result is the bump observed in
Fig.~(\ref{fig7}.B) at low force ratios: here not any of the
effective-charge models, but the Yukawa pair-force model as such fails
dramatically, predicting a total force that is five times smaller in
magnitude than the correct many-body force. We have investigated the
local structure around the colloids feeling these forces. They are
always located in regions of larger local density. Here, the local
mean-distance between the colloidal particles is relatively short
(compared to the suspension-wide mean distance) and the Yukawa pair
potential is thus probed at rather short inter-particle
distance. However, at these distances expression (\ref{eq7}) is bound
to fail, as we have indicated already in sec.~(\ref{sec:2models}).
The extent of this failure is now quantified in Fig.~(\ref{fig7}.B).

\section{Test II: Thermodynamic consistency and the Kirkwood Buff-relation}
\label{sec:structthermo}

Having seen in the previous section that the distribution of forces
felt {\it in situ} by the colloids in the mixture can in most cases be
captured by a Yukawa effective pair potential, we now turn to the
second test of these potentials and relate the compressibility of the
suspension to its structure as computed within the OCM. The way to do
so is to apply the Kirkwood-Buff relation introduced further below.

$\Pi_{\text{micro}}$ of both the PB-cell and the Jellium model is
depicted in Fig.~(\ref{fig2}). It provides an excellent approximation
to the total pressure of salt-free suspensions: both models lead to a
pressure that is in very good agreement with existing experimental
data \cite{reus} and primitive model simulations \cite{linse00,rque4},
see e.g.  \cite{LevinJPCM,trizaclevin,nota}. The Kirkwood-Buff
relation (\ref{eq:kirkwood}) now allows for a test of the effective
potential chosen: $u_{\text{eff}}$ should lead to a colloid structure,
embodied in a long-wavelength structure factor $S(0)$ computed within
the OCM, that is compatible with the compressibility following from
$\Pi_{\text{micro}}$.  As the same effective parameters appear both in
the Yukawa-forces leading to $S(0)$ and in $\Pi_{\text{micro}}$ of
eq.~(\ref{eq14}) this provides a critical consistency test of our
effective-charge models.

We will also report results at medium and high salt concentrations,
where it does not seem possible to test the effective potential as
severely as in the no salt case, but where it is still of interest to
compare the different pressures introduced in this section.  A similar
idea has previously been worked out by Lobaskin et al. \cite{vladimir}
who checked the consistency of thermodynamic and structural properties
of an asymmetric electrolyte containing macroions with 60 elementary
charges and monovalent counterions.  Primitive-model results have been
compared with different more approximate theories such as the cell
model (within mean-field and even beyond). Finally, technical details
may be found in Appendix \ref{app:tech} and \ref{app:integral}.

We start this section with an outline of the theoretical background
and present the actual results of our second test in
section~\ref{sec:KB}.

\subsection{Theoretical background}
\label{sec:theo}

\subsubsection{The osmotic pressure of a charge-stabilized colloidal suspension}

Our problem is best treated in a semi-grand canonical ensemble.  $N_c$
colloids are confined to a certain volume $V$ which the small solvent
molecules and additional micro-ions are free to leave; they are
coupled to a reservoir fixing their chemical potentials. For the
moment, we need not distinguish between the different types of small
molecules, but may denote them collectively by an index $s$, while the
colloidal degrees of freedom are referred to by the index $c$. The
total internal energy of the system can then be written as
$U=U_{cc}+U_{ss}+U_{cs}$ where the index combinations indicate over
what degrees of freedom the bare interactions have to be summed. The
force on particle $i$ ($i \in c,s$) then is ${\bm F}_i = - {\bm
\nabla}_i U$ and the total pressure of the colloidal suspension can be
obtained using the virial
\begin{equation}  \label{eq0}
P = \rho kT  + \frac{1}{3V}  
\left\langle \sum_{i \in c,s} {\bm r}_i \!\cdot\! {\bm F}_i \right\rangle_{s,c}
\end{equation}
where the thermal average has to be taken over the colloidal as well
as solvent/solute degrees of freedom. The first term, the ideal-gas
term, is just the sum of densities of all components.

Within the OCM, the pressure reads
\begin{equation}  \label{eq3}
\Pocm = \rho_c kT  + 
\frac{1}{3V}  \left\langle \sum_{i \in c} {\bm r}_i \!\cdot\! {\bm F}_i^{\text{eff}} 
\right\rangle_{c}
\end{equation}
where the forces are obtained from the gradient of the effective
potential and the summation runs over the $N_c$ colloids only.

By definition, the effective potential $u_{\text{eff}}$ reproduces the
same pair (colloid-colloid) distribution function $g(r)$ as that of
the original mixture, assuming a pair-wise Hamiltonian, see
e.g. \cite{luc}. However, there is no guarantee that the corresponding
sum of effective pair forces coincides, for a given colloid $i$ in a
given colloidal configuration, with the micro-ion average of $-{\bm
\nabla}_i U$ which provides the true force.  The comparison between
both forces has been made in section \ref{sec:manybody}. For the sake of
the discussion, we assume here that they coincide.  Eq.~(\ref{eq3})
then uncovers one severe deficiency of the OCM picture : $\Pocm$ only
provides a contribution to the total pressure given by
Eq. (\ref{eq0}). In particular, in the limit of low salt the
micro-ions are known to dominate the overall pressure of the
suspension: that is the reason why $\Pi_{\text{micro}}$ of
Fig.~(\ref{fig2}) provides such an excellent approximation to the
total pressure. This implies on the other hand that $\Pocm$ is a very
poor approximation for $P$. The density derivatives of both pressures
are however intimately connected, under all conditions, as we discuss
in the following section (see eq.~(\ref{eq:Pocmcomp})).

\subsubsection{The Kirkwood-Buff relation}

The Henderson theorem \cite{hend74} guarantees that every
pair-correlation function $g(r)$ can be uniquely associated with one
pair-potential. As for the system considered here, this implies that
for any given colloid density, there exists in principle one, and only
one, effective pair-potential leading to a $g(r)$ within the OCM that
is in perfect agreement with the correct colloid-colloid pair
correlation of the full multi-component system. Unfortunately, this
does not guarantee that in such a multi-component system the
higher-order correlation functions are equally well reproduced (see
for example \cite{russ05} where pair-, but not triplet correlations of
a many-component colloidal systems could be reproduced within a simple
pair-interaction picture).

However, there is one important quantity one can still derive if the
correct colloid-colloid $g(r)$ is known: the osmotic isothermal
compressibility $\chi$ defined in eq.~(\ref{eq4}). The Kirkwood-Buff
relation relates this thermodynamic quantity to the infinite
wavelength limit of the colloid-colloid structure factor $S(q)$
\cite{Kirkwood}
\begin{equation}  
\label{eq:kirkwood}
\chi \, = \frac{\partial \rho_c}{\partial \beta \Pi}
\biggl|_{T,\text{salt}}  
\, = \,S(0)
\end{equation}  
where $S(0)$ is related to $g(r)$ via
\begin{equation}  \label{eq6}
S(0)=1+\rho_c \int (g(r)-1) d{\bm r} \:.
\end{equation}  
Remarkably, Eq.~(\ref{eq:kirkwood}) is exact: nothing about the
microion-microion correlations or the microion-colloid correlations
needs to be known, but only the colloid-colloid pair correlations are
required to compute the correct osmotic compressibility of the full
multi-component system. By contrast, the full compressibility of the
system, defined as $\chi^{-1}=\partial \beta P / \partial
(\rho_s+\rho_c)$ with $P$ from Eq.~(\ref{eq0}), depends on the pair
correlations of all components \cite{luc}.

Connecting thermodynamic to structural information,
Eq.~(\ref{eq:kirkwood}) is of central importance in the present work.
In many colloidal suspensions the micro-ions determine the
thermodynamics, while the macro-ionic degrees of freedom are more
important for the structural properties of the system.  In these cases
Eq.~(\ref{eq:kirkwood}) is also well suited to bridge the gap between
the micro-ion and the macro-ion oriented viewpoints, thus providing a
severe test for the quality of the effective potential. Indeed, though
having a clear-cut definition, an effective potential is nevertheless
a difficult object to compute and one often postulates its functional
form, just as we have done in eq.~(\ref{eq7}). With
eq.~(\ref{eq:kirkwood}) we can investigate {\it a posteriori} the
relevance of the underlying {\em ad-hoc} assumptions made in deriving
the effective potentials.  Regarding the effective-charge concepts
under scrutiny here, eq.~(\ref{eq:kirkwood}) describes a route how to
connect $u_{\text{eff}}$ with $\Pi_{\text{micro}}$.  As the same
effective parameters appear in both quantities, this relation also
provides a consistency test of our effective-charge models.

\subsubsection{On the various contributions to the total pressure}

So far we have introduced three different pressures: $P$ from which
the osmotic pressure $\Pi$ follows, $\Pocm$, and finally $\Pmicro$,
the latter quantity being for instance the pressure found in the cell
or in the Jellium models, see eq.~(\ref{eq14}).  At this point, it
seems worth discussing the non-trivial connection between these three
pressure expressions, at the expense of introducing a fourth quantity:
we define a ``colloidal'' pressure $\Pcoll$ from the difference
between the total pressure $P$ and $\Pmicro$ \be P = \Pcoll + \Pmicro.
\label{eq:Pmicro}
\ee
The pressure $\Pmicro$ can be considered as arising from so-called
``volume terms'' in the total free energy of the system, see e.g.
\cite{Rene,Denton}.

In the situation where the system is confined in a {\em closed} 
box of volume $V$
($\rho_c=N_c/V$), 
it can be shown that \cite{rque19,rque20}
\begin{equation}
P \,=\, \rho_c \,kT +  \frac{1}{3V}
\left\langle\sum_{i\in c} {\bm r}_i \!\cdot\! {\bm F}_i^{\text{eff}} 
\right\rangle +
\frac{kT}{3V} \left\langle\oint_{\text{box}}
\rho_{\text{micro}}({\bm r}) \, 
{\bm r}\!\cdot\! d^2{\bm S}\right\rangle,
\label{eq:65}
\end{equation}
The third term on the right hand side in (\ref{eq:65}), for which the
surface integral with normal oriented outward runs over the box
confining the system, accounts for the direct coupling between
colloids and micro-ions. It is precisely this quantity that the
micro-ionic cell-model approaches aim at computing. In other words,
this third contribution may be identified with $\Pmicro$ in
(\ref{eq:Pmicro}) so that remembering Eq. (\ref{eq3}), we obtain
$\Pcoll \simeq \Pocm$. However, a closed cell is not the most
convenient configuration (in particular, the effective potential of
interaction not only depends on the relative distance between two
colloids, but also on the distance to the confining walls), and from a
computational perspective, it is often more desirable to work with
periodic boundary conditions systems. Unfortunately, Eq. (\ref{eq:65})
which provides a physically transparent interpretation to $\Pcoll$ in
the confined case (close cell), breaks down with periodic boundary
conditions.  This failure is discussed at length in \cite{submitted},
together with the fact that in low salt (or no salt) conditions, one
has $\Pcoll \ll \Pmicro$ which implies $P\simeq \Pmicro$ (see
\cite{ecc} for a related discussion).

From the previous discussion, it appears that the connection between
$\Pocm$ and $P$ is not straightforward. The simplest relation between
both quantities follows from the remark that the colloidal structure
within the OCM is of course the same as in the original mixture
(assuming the effective potential to be the correct one).
Eqs.(\ref{eq:kirkwood}) and (\ref{eq6}) then dictate that the
compressibilities in both approaches coincide : \be \frac{\partial P
}{\partial \rho_c}\biggl|_{T,\text{salt}} \,=\, \frac{\partial
\Pocm}{\partial \rho_c}\biggl|_{T,\text{potential}}
\label{eq:Pocmcomp}
\ee where it is crucial to compute the OCM compressibility {\em at
constant potential of interaction $u_{\text{eff}}$}, i.e. discarding
any density dependence of the effective potential. In a region of
parameter space where the density dependence of $u_{\text{eff}}$ is
absent or weak enough, the ``salt'' and ``potential'' subscripts in
the partial derivatives of Eq. (\ref{eq:Pocmcomp}) correspond to the
same constraint, since $u_{\text{eff}}$ then only depends on the salt
chemical potential, besides relative distance.  It is then possible to
integrate Eq. (\ref{eq:Pocmcomp}) to obtain $P\simeq \Pocm$.  Such a
situation is met in the salt dominated regime to be discussed below.

\subsection{Applying the Kirkwood-Buff relation}
\label{sec:KB}

We next apply eq.~(\ref{eq:kirkwood}) to the case at hand and expect
to find the following: In the no salt case where the total pressure is
very close to $\Pmicro$ we should obtain within the OCM a
long-wavelength structure factor $S(0)$ compatible with $\partial
\rho_c/\partial (\beta \Pmicro)$. Any difference between both
quantities indicates a lack of consistency in the (somewhat
uncontrolled) procedure leading to the effective Yukawa potential
(\ref{eq7}) chosen here.  Such an inconsistency may take its roots in
an incorrect computation of effective parameters (charge and screening
length) or more fundamentally in the functional screened Coulomb form
taken.

\begin{figure}
\includegraphics[width=0.66 \textwidth]{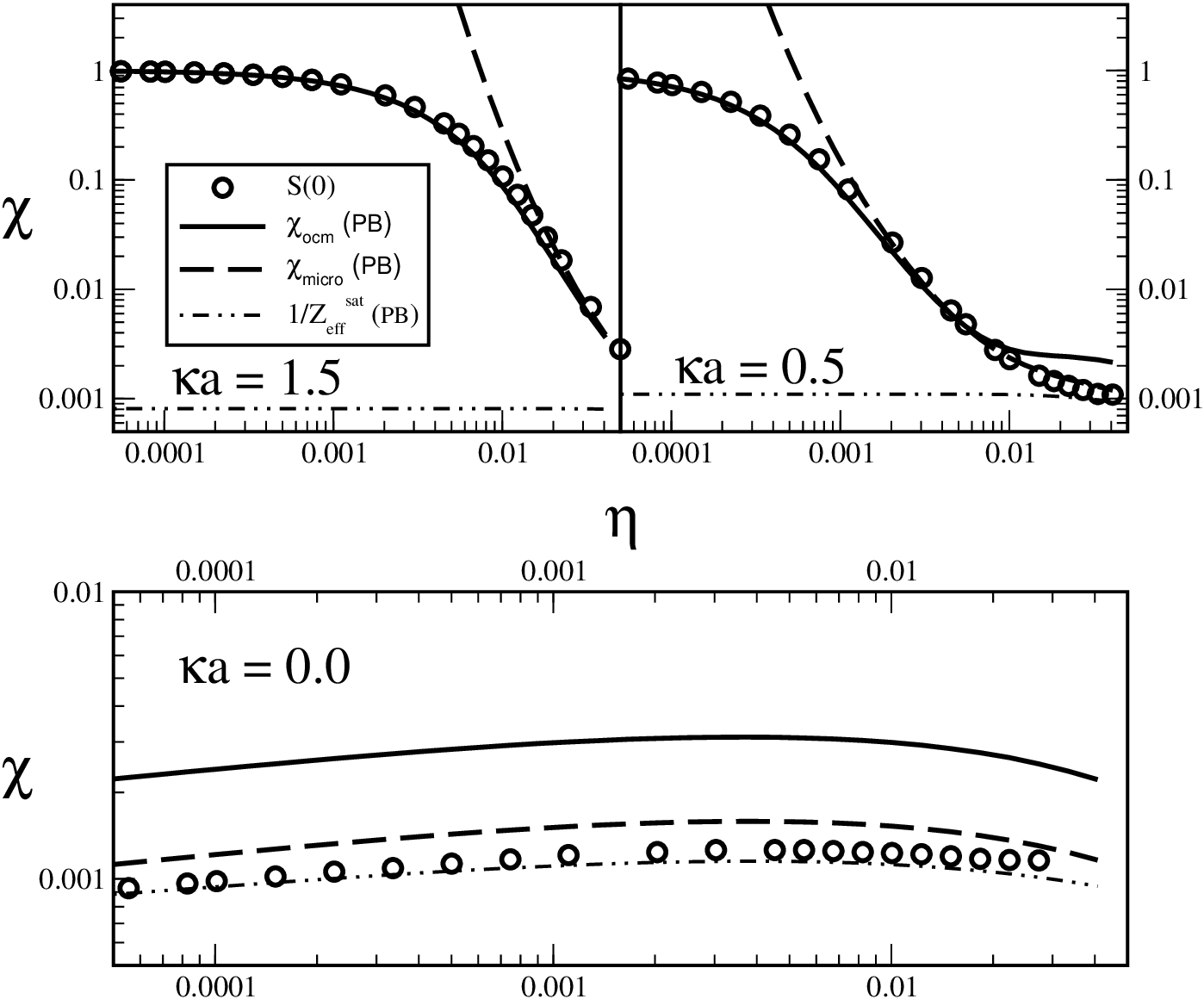}
\caption{\label{fig4} 
Compressibility vs volume fraction, for different salt conditions and
applying different approximations as discussed in the text.  Dashed
line: $\chi_{\text{micro}}(\rm PB)$, solid line:
$\chi_{\text{ocm}}(\rm PB)$ and empty circles: the colloid-colloid
$S(0)$ from Fig.~(\ref{fig3}).  Dashed-dotted line shows
$1/Z_{\text{eff}}^{\text sat} (\rm PB)$, which is a valid approximation at all
densities without salt, and holds at high densities only, when salt is
present.}
\end{figure}
\begin{figure}
\includegraphics[width=0.66 \textwidth]{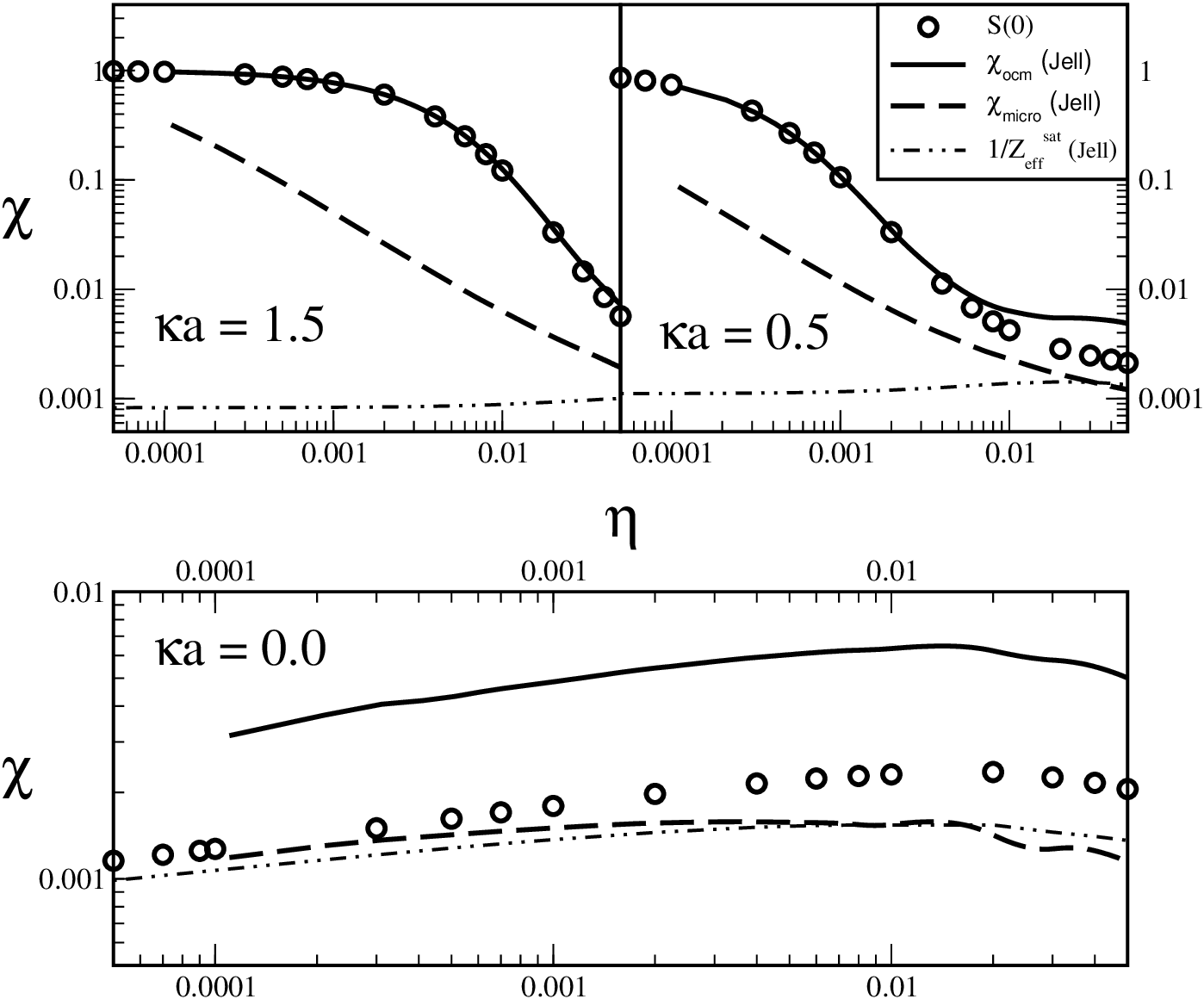}
\caption{\label{fig5} Same as Fig.~(\ref{fig4}), but with 
effective parameters derived from the Jellium model.
Dashed line:  $\chi_{\text{micro}}(\rm Jell)$, 
solid line:  $\chi_{\text{ocm}}(\rm Jell)$, empty  circles: $S(0)$, 
dashed-dotted line: $1/Z_{\text{eff}}^{\text sat} (\rm Jell)$ }
\end{figure}

We have calculated $S(0)$ within the OCM as a function of $\eta$ as
described in detail in the appendix section \ref{app:tech}
(Monte-Carlo simulation) and \ref{app:integral} (integral-equation
theory). Both methods lead to identical results, shown in
Fig.~(\ref{fig3}) of the appendix. Such sets of $S(0)$ curves had to
be computed twice, using once the PB cell model parameter and once the
Jellium model parameter within the effective Yukawa pair-potential.
Figures \ref{fig4} (PB-cell) and \ref{fig5} (Jellium) show the
corresponding $S(0)$ curves. Each figure consists of three graphs,
corresponding to $\kappa a =1.5$, $\kappa a =0.5$ and $\kappa a =0.0$.
In each graph, we compare $S(0)$ to the compressibility
$\chi_{\text{micro}}$ following from the $\rho_c$ derivative of
pressure $\Pmicro$ (or equivalently $\Pi_{\text{micro}}$). To
distinguish between the two effective charge models used to compute
$\Pi_{\text{micro}}$ via eq.~(\ref{eq14}), we introduce the notation
$\chi_{\text{micro}}(\rm PB)$ and $\chi_{\text{micro}}(\rm Jell)$.
Alternatively, one may compute the OCM pressure $\Pocm$, defined in
(\ref{eq3}), and obtains then the compressibility $\chi_{\text{ocm}}$
as the $\rho_c$ derivative of $\Pocm$. If PB-cell model (Jellium
model) parameters have been used in the effective Yukawa potential, we
denote the resulting compressibility by $\chi_{\text{ocm}}(\rm PB)$
($\chi_{\text{ocm}}(\rm Jell)$). These quantities, and in addition the
inverse of the saturated effective charges, are also graphed in Figure
\ref{fig4} and \ref{fig5}.

\subsubsection{No salt}

We observe that $\chi_{\text{micro}}$, in the no salt
case, is close to $S(0)$. The agreement seems to be slightly better
for Jellium compared to cell parameters at low density, while the
opposite holds at higher densities. This finding is consistent with
the fact that the probability distribution shown in Fig. \ref{fig7}.B
(where $\eta=10^{-4}$, a low value) is slightly more peaked for the
Jellium than for the cell.  On the other hand, the inadequacy of
$\chi_{\text{ocm}}$ for $\kappa a=0$ is expected : $\Pocm$ has here
nothing to do with the total pressure $P$ ; the resulting
compressibility is about a factor of two larger than $S(0)$. This
question is addressed in \cite{submitted} and for completeness, we
adapt the argument in Appendix \ref{app:ocmpress}.  As can be seen in
Figs. \ref{fig4} and \ref{fig5}, the no-salt compressibility is close
to the inverse effective charge.  This stems from the fact that at
least within the Jellium, one has $\beta \Pmicro = \Zeff \rho_c$ for
$c_s=0$ and since the density dependence of $\Zeff$ is mild
(logarithmic, see \cite{trizaclevin}), one has
$\chi_{\text{micro}}\simeq 1/\Zeff$. Within the cell model, $\beta
\Pmicro = \Zeff \rho_c$ is no longer exact, but accurate enough to
lead again to $\chi_{\text{micro}}\simeq 1/\Zeff$.

\subsubsection{Added salt}

In the first two figures in Figs.\ref{fig4} and \ref{fig5} (the ones
with $\kappa a>0$) we can see two regimes. At low volume fraction the
OCM compressibility $\chi_{\text{ocm}}$ is a good approximation and
the micro-ionic $\chi_{\text{micro}}$ fails, while at high enough
volume fraction the reverse is true, $\chi_{\text{micro}}$ is good and
$\chi_{\text{ocm}}$ fails. 

The derivative involved in the computation of $\chi_{\text{ocm}}$
includes the density dependence of the effective parameters.  There
are however situations where these parameters are virtually
independent of $\rho_c$ (see Fig. \ref{fig1}). This is the case for
example at low density or high enough salt concentration.  Looking
back at Eq.  (\ref{eq:Pocmcomp}), one realizes that in these cases the
constant potential of interaction constraint coincides with that of
constant $c_s$. Then Eq.~(\ref{eq:Pocmcomp}) dictates that
$\chi_{\text{ocm}}$ {\it must} be equal to $S(0)$. It is therefore
not a surprise, nor a deep finding, that in Figs.  \ref{fig4} and
\ref{fig5} the $S(0)$ coincides with $\chi_{\text{ocm}}$ in all those
situations where the effective parameters have no $\rho_c$ dependence,
i.e., at high enough salt concentration and at low density.

Such an agreement notwithstanding allows us to compute the full
pressure $P$ when the effective potential is density independent.
Remembering that at high salt content, the effective forces following
from $u_{\text{eff}}$ are very close to their ``exact'' counterpart
(see Fig. \ref{fig7}), we expect $\Pocm$ to provide there a good
approximation to $P$, at least at not too high densities, i.e.  in the
salt-ion dominated screening regime when $\eta \ll \eta^*$ with
$\eta^*$ from eq.~(\ref{eq:etastar}).

At high $\eta$ the counter-ions dominate the screening. Thus, no
matter how large $\kappa a$ is, there is always a regime at high
enough $\eta$ where the salt-ions can be neglected and where the
results become independent of $\kappa a$. That implies that curves
differing in $\kappa a$ must approach the same value at high enough
$\eta$. That can be observed, for example, in Fig.~\ref{fig1}.A and
Fig.~\ref{fig3} of the appendix, where all curves show the same high
$\eta$ behavior.  The same applies to Figures \ref{fig4}
and \ref{fig5}: At high densities, all curves approach the zero-salt
curves and go to $1/Z_{\text{eff}}$, which is the same for all $\kappa
a$ in this counter-ion dominated limit. On the other hand, for $\eta
\to 0$ the compressibilities for $\kappa a \neq 0$ approach the
ideal-gas limit.

We have already remarked that for $\kappa a=0$, $\chi_{\text{micro}}$
coincides with the OCM $S(0)$. Since, for large $\eta$ (i.e.
$\eta>\eta^*$) all $\kappa a\neq 0$ curves must approach the $\kappa
a=0$ curve, we expect to find that at high enough $\eta$,
$\chi_{\text{micro}}$ coincides with $S(0)$, since then the system is
close to the salt-free regime \cite{rque22}.  This is indeed the case
in Fig \ref{fig4} (and to a lesser extent in Fig. ~\ref{fig5} where
only the trend is visible). This limiting behaviour seems to
be a natural self-consistency requirement to impose to any effective
potential.
\begin{figure}
\mbox{}\\[1cm]
\includegraphics[width=0.66 \textwidth]{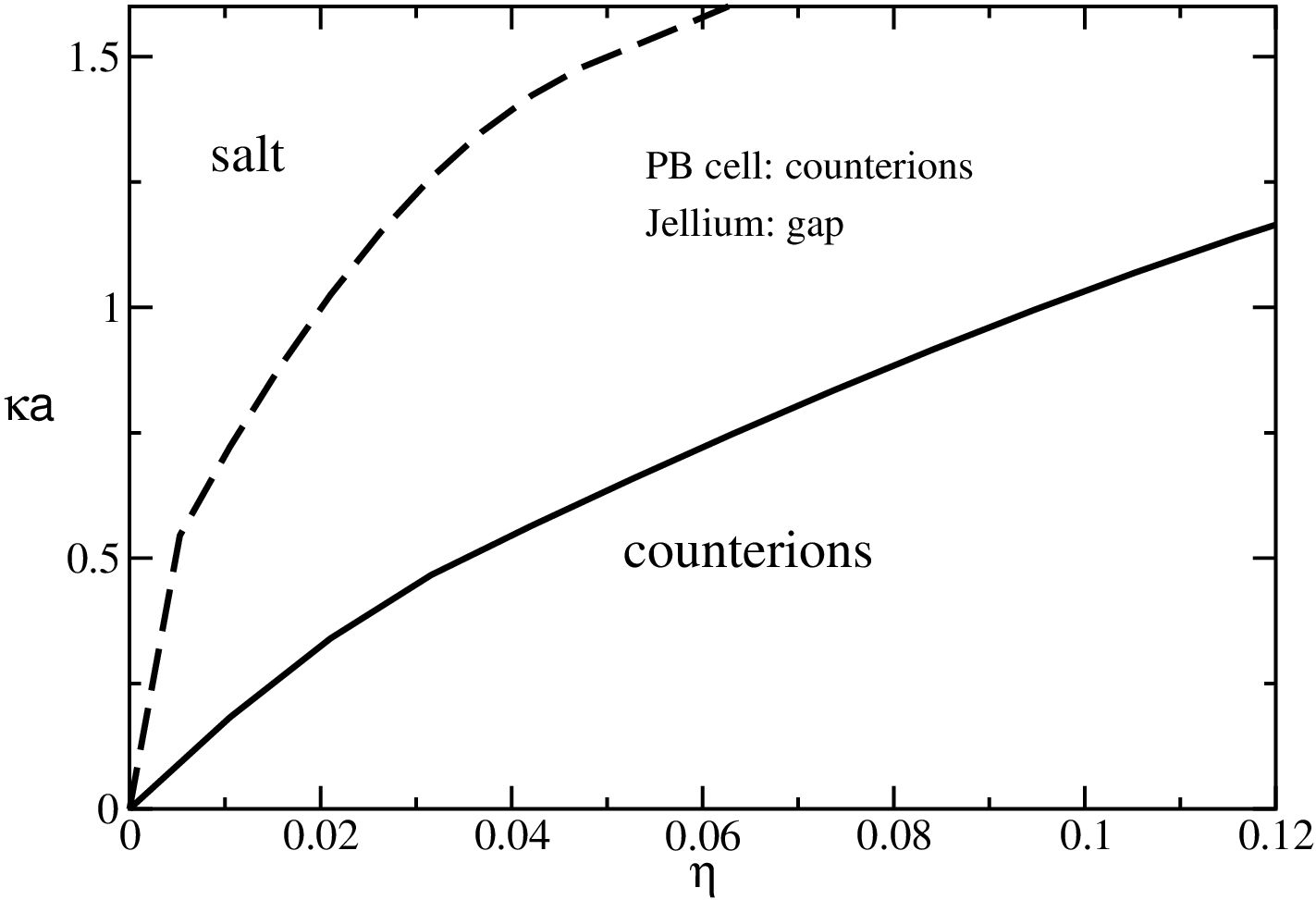}
\caption{\label{Pfig} A diagram showing the regions in parameter space
of salt dominated screening and counter-ion dominated screening for
both models.  In the salt dominated screening regime (left from the dashed line) the OCM compressibility $\chi_{\text{ocm}}$ is consistent with $S(0)$, while in the counterion dominated screening regime (right from the dashed line for the cell model and right from the solid line for the Jellium) $\chi_{\text{micro}}$ is consistent with $S(0)$. On the solid line, $\eta$ is equal to $\eta^*$ defined in eq.~(\ref{eq:etastar}), while on the dashed line $\eta$ is equal to $0.2 \eta^*$.}
\end{figure}

For the Jellium model we have already defined the crossover volume
fraction $\eta^*$ (eq.~(\ref{eq:etastar})) separating the salt and
counterion dominated screening. The dependence of $\eta^*$ on the
salt concentration $\kappa a$ and on the volume fraction $\eta$ is
graphed in Fig.~\ref{Pfig} (solid line). This graph is meant to
summarize our findings of Fig. \ref{fig4} and \ref{fig5}.  For the
Jellium model, the counterion dominated regime ($\eta>\eta*$) is to
the right of the solid line. Here, the OCM $S(0)$ is found from
Fig. \ref{fig5} to be consistent with $\chi_{\text{micro}}$.  The
other regime (the salt-ion dominated regime) where we can observe from
Fig. \ref{fig5} that $S(0)$ is consistent with $\chi_{\text{OCM}}$ is
unfortunately {\em not} located to the left of the solid curve, but
rather to the left of the dashed line empirically determined as
$\eta^*_{s}=0.2\eta^*$. This means that in frame of the Jellium model 
there is a gap between the salt dominated region and counterion 
dominated region ($\eta^*_s < \eta < \eta^*$). At these intermediate 
volume fractions neither $\chi_{\text{micro}}(\rm Jell)$ nor 
$\chi_{\text{ocm}}(\rm Jell)$ is consistent with $S(0)$.

Such a gap is not found using parameters from the PB-cell model. For
the PB cell model we have no analytical prediction for the value of
the crossover volume fraction and have to read the values from the results in
Fig. \ref{fig4}. Empirically we have found the crossover volume fraction of about 
$\eta\approx 0.2\eta^*$, which is the same as the previously defined $\eta^*_s$ for
the Jellium model. This is plotted as a dashed line in Fig.\ref{Pfig}.
For the PB-cell model, we may then summarize Fig. \ref{fig4} as
follows.  For systems parameters lying on the right hand side of the
dashed curve, $S(0)$ is consistent with $\chi_{\text{micro}}(\rm PB)$,
for those system parameters on the left hand side $S(0)$ is consistent
with $\chi_{\text{OCM}}(\rm PB)$.

We emphasize at this point that with salt at high $\eta$, a better
self-consistency is not necessarily synonymous with the fact that
$\Pmicro$ is in itself a better approximation for $P$. The situation
is different for $\kappa a=0$ where we have the extra piece of
knowledge that $P\simeq \Pmicro$, that comes from comparison with
experiments or primitive model computations (see also section
\ref{sec:primitive}).

After this discussion, we can summarize the results of our
Kirkwood-Buff consistency check of the effective Yukawa
pair-potentials and the two effective charge models, as follows:
\begin{itemize}
\item For $\kappa a =0$, we find consistency essentially for both
effective-charge models where the agreement is better for the Jellium
model at low densities and for the PB-cell model at high densities.
\item For $\kappa a =0$, the compressibility derived from $\Pocm$ is
inadequate and not compatible with the OCM $S(0)$. The OCM approach
with the Yukawa potential as effective pair-potential seems to produce
correct results for the structure of the suspension, but not for the
pressure.
\item
The no-salt compressibility is close to the inverse effective charge.
This offers a convenient way to estimate compressibilities
$\chi_{\text{micro}}$ at $\kappa a =0$.
\item For $\kappa a \neq 0$ and low volume fraction ($\eta < \eta_s^*$),
consistency is found between $\chi_{\text{ocm}}$ and $S(0)$, but not
between $\chi_{\text{micro}}$ and $S(0)$. The OCM approach makes sense
for calculating both the structure and the pressure.
\item For $\kappa a \neq 0$ and high volume fraction ($\eta > \eta^*$
for Jellium, $\eta>\eta^*_s$ for PB-cell), we get back to the
zero-salt case where $S(0)$ is consistent with $\chi_{\text{micro}}$,
but not with $\chi_{\text{ocm}}$.
\item The value of $\eta^*_s=0.2\eta^*$ has been empirically determined at the constant value of $\lambda_B/a=0.01$. This result might depend on the value of $\lambda_B/a$, which has not been studied in scope of this paper.
\end{itemize}

\begin{figure}
\mbox{}\\[1cm]
\includegraphics[width=0.46 \textwidth]{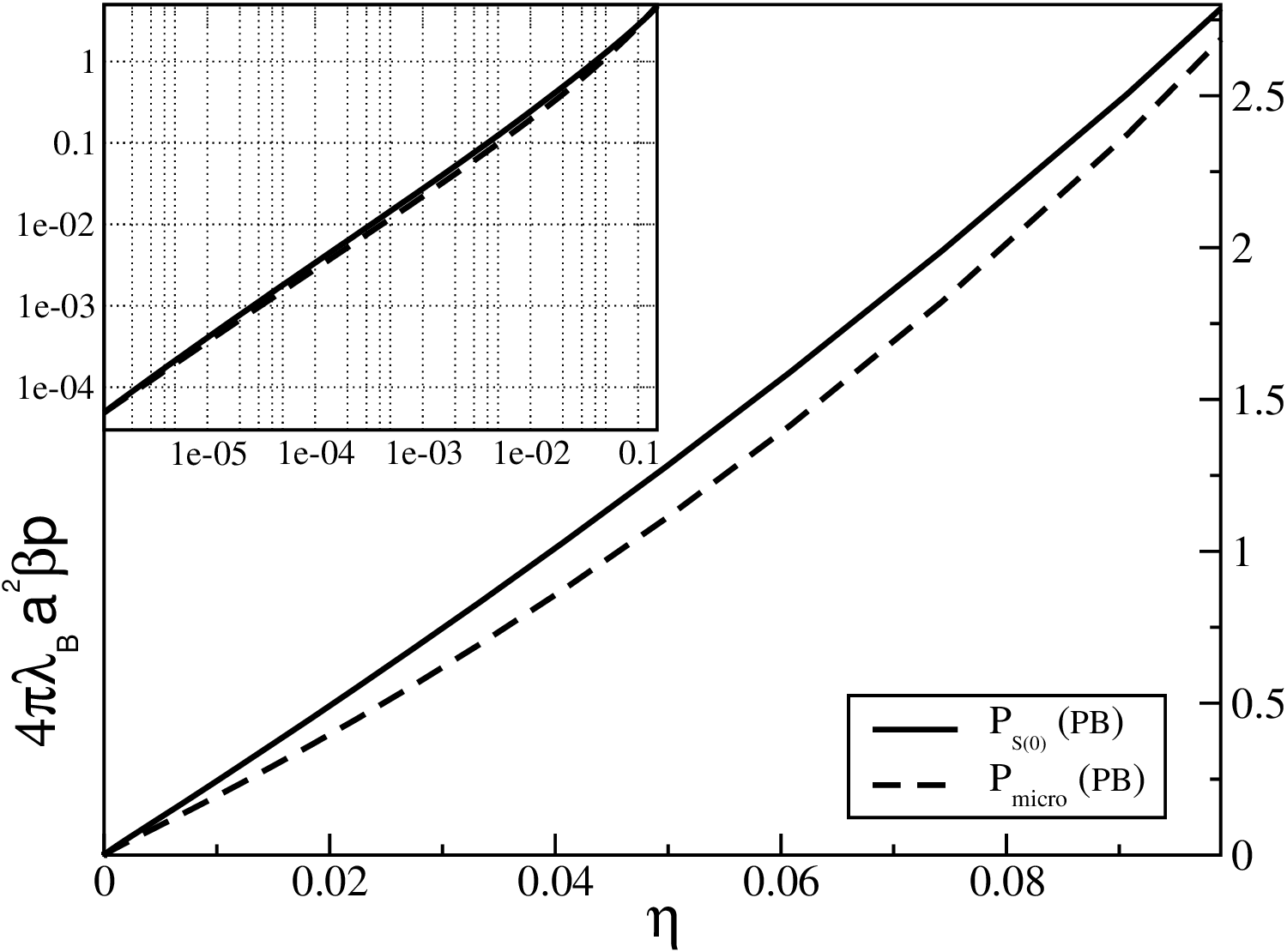}
\hfill
\includegraphics[width=0.46 \textwidth]{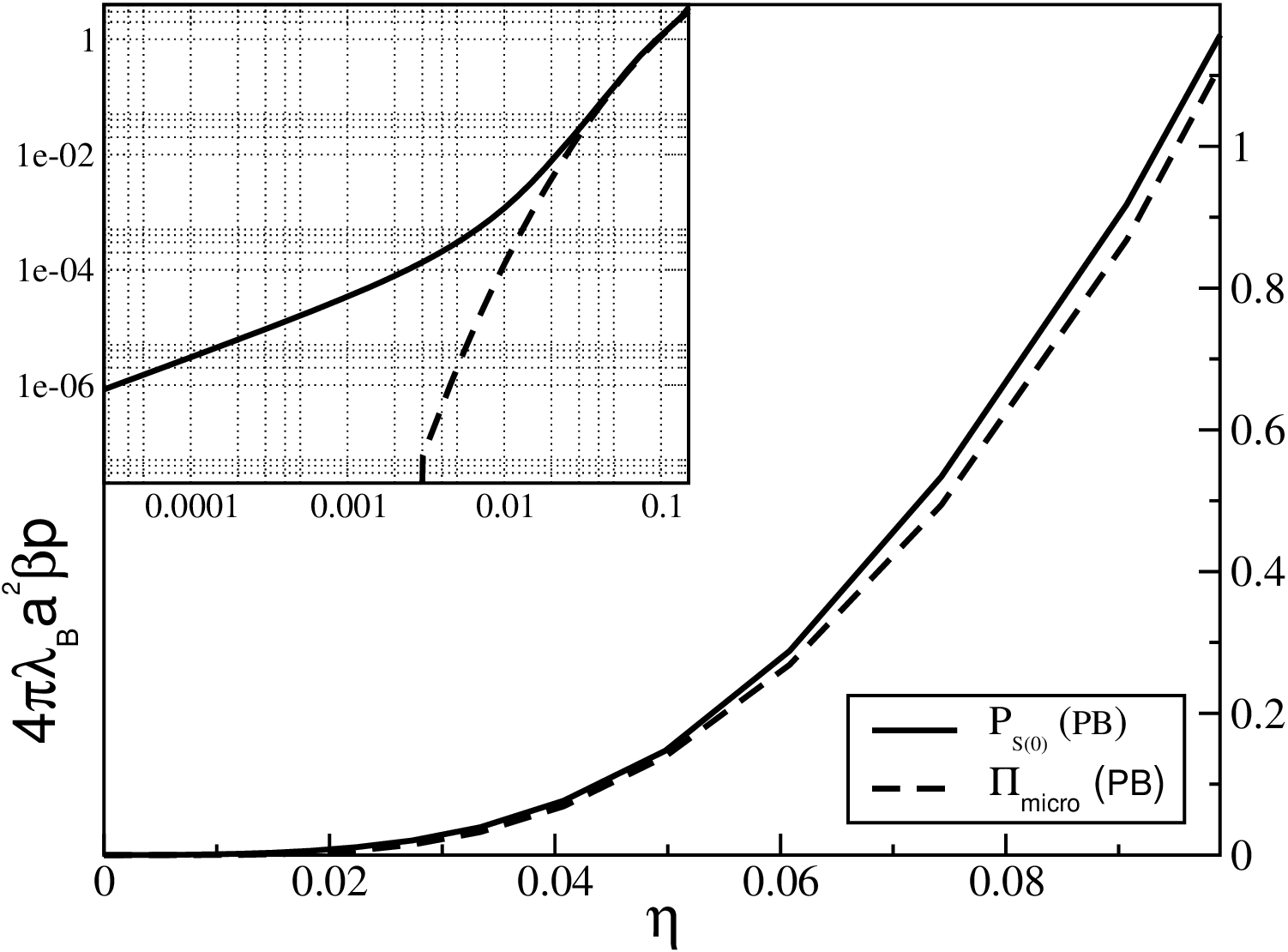}
\caption{Comparison between the PB cell model pressure and the
associated $P_{S(0)}$ (computed within the OCM with PB cell effective
parameters in $u_{\text{eff}}$), for $\kappa a=0$ (left figure) and
$\kappa a=1.5$ (right figure).}
\label{Pr00}
\end{figure}

In cases where we have reason to assume that the OCM $S(0)$ is close
to the correct $\chi$, we can of course estimate the pressure
$P_{S(0)}$ from integrating the OCM $S(0)$
\begin{equation}
\beta P_{S(0)}(\rho) \,=\, \int_0^{\rho} \frac{1}{S(0)(\rho')} \,d\rho'\:.
\label{eq:PS0}
\end{equation}
To illustrate this idea, we have performed this integration in
Fig.~\ref{Pr00} for $\kappa a=0$ and $\kappa a=1.5$, using the PB-cell
effective parameters. The resulting curves can then be compared to
$\Pi_{\text{micro}}$ from eq.~(\ref{eq14}) (again with the PB-cell
parameters), which for $\kappa a=0$ we know to be an excellent
approximation to the full pressure of the suspension.  We have data
for $S(0)$ starting at volume fraction $\eta_0\approx 10^{-8}$, which
is used as the lower bound of integration in Eq. (\ref{eq:PS0}). A
small constant is then added to the right hand side of (\ref{eq:PS0}),
so that the result coincides with $\Pi_{\text{micro}}$ at $\eta_0$.
The figure demonstrates that $P_{S(0)}$ agrees well with
$\Pi_{\text{micro}}$, surprisingly, not only for $\kappa a=0$ but also
for $\kappa a=1.5$. However, at very low $\eta$, the agreement is
excellent for $\kappa a=0$, but not for $\kappa a=1.5$, as one would
expect from Fig.~(\ref{fig4}).  Note that a hypothetical discrepancy
between $\Pmicro$ and $P_{S(0)}$ here could not be considered as a
lack of self consistency, since the contribution $\Pcoll$ in
$P=\Pcoll+\Pmicro$ can be non-negligible.

%%%%%%%%%%%%%%%
\section{Test III: Comparison to primitive model data}
\label{sec:primitive}

\begin{figure}
\mbox{}\\[1.0cm]
\includegraphics[width=0.46 \textwidth]{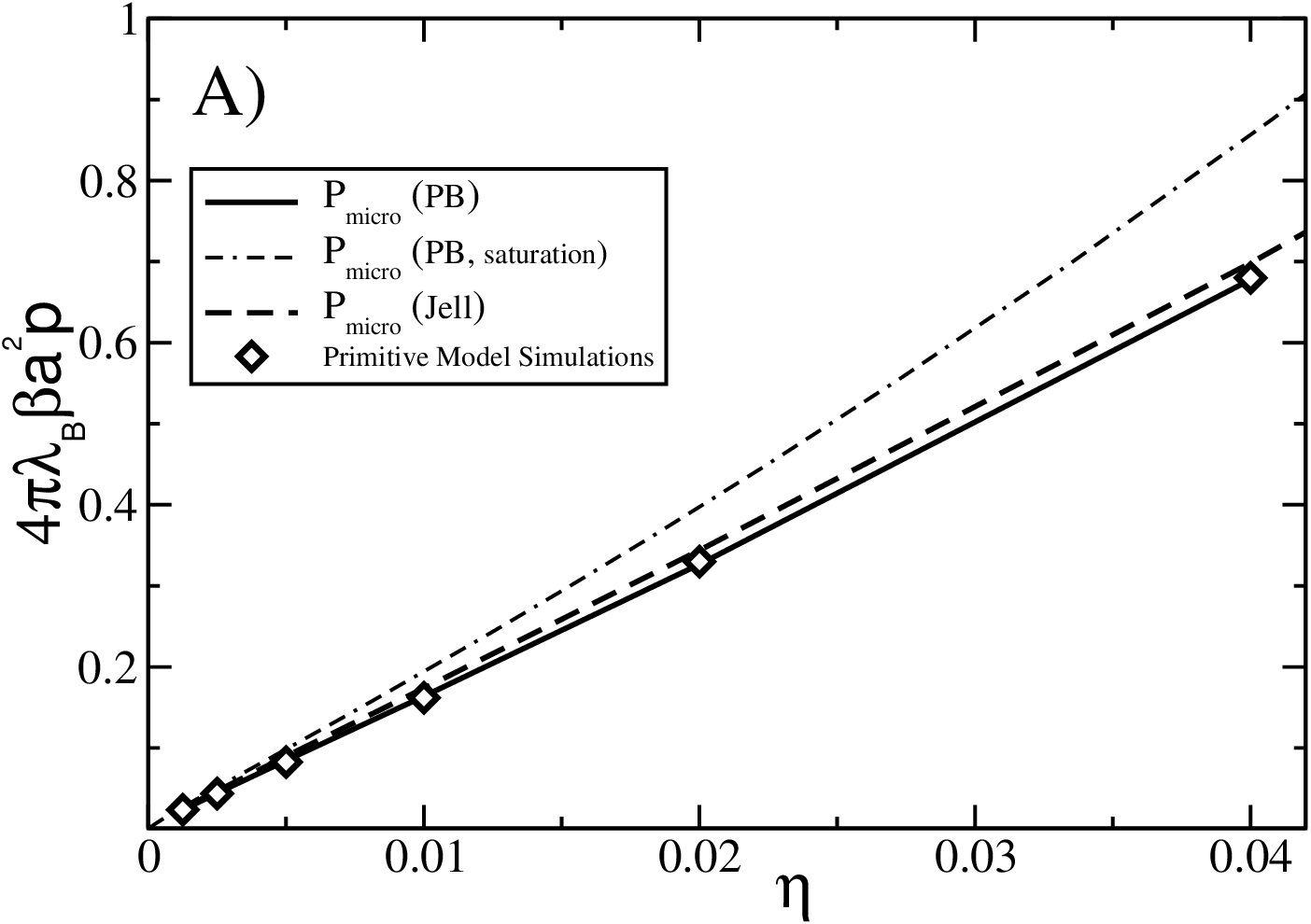}
\hfill\includegraphics[width=0.46 \textwidth]{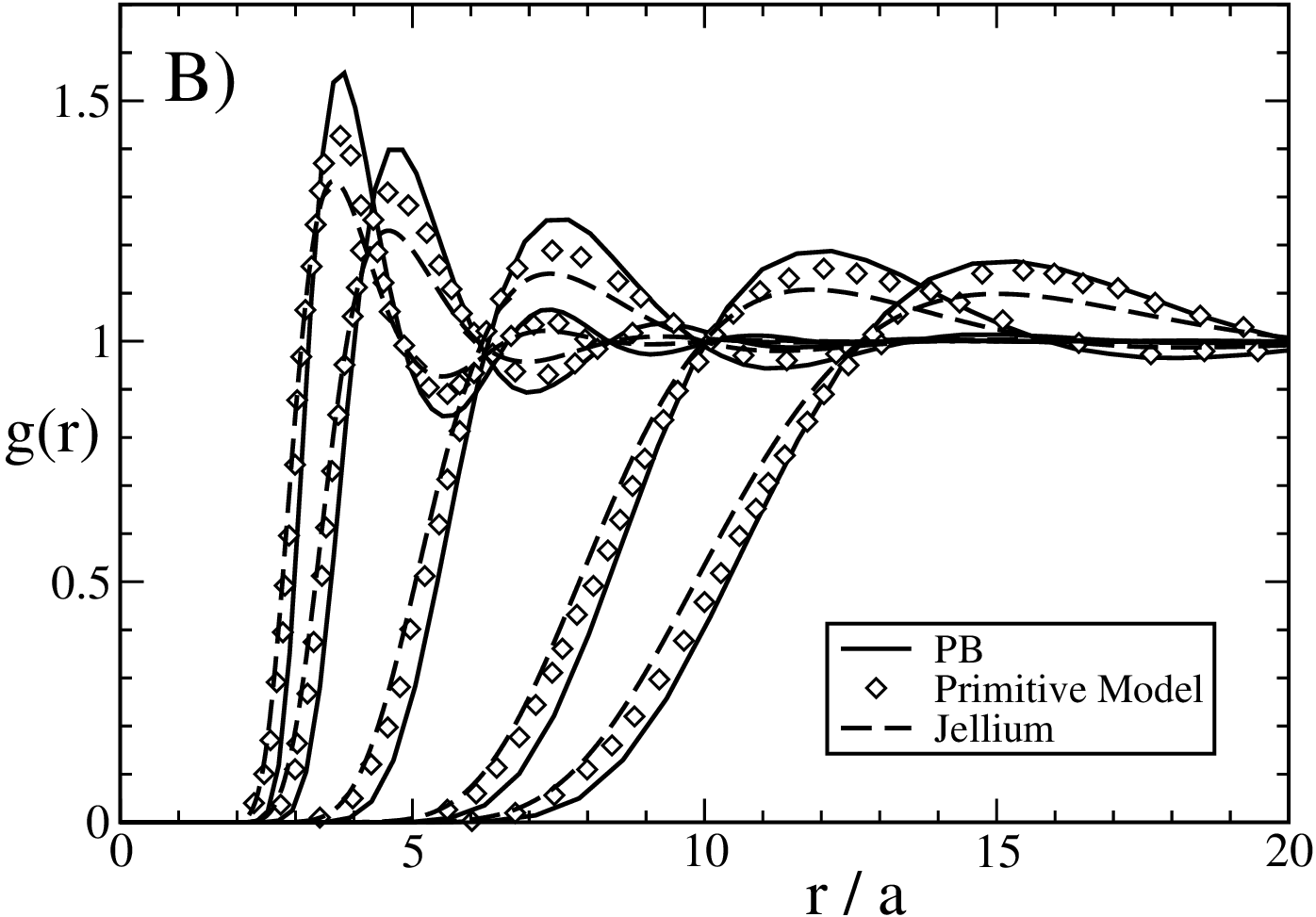}
\mbox{}\\[2.750cm]
\centering\includegraphics[width=0.66 \textwidth]{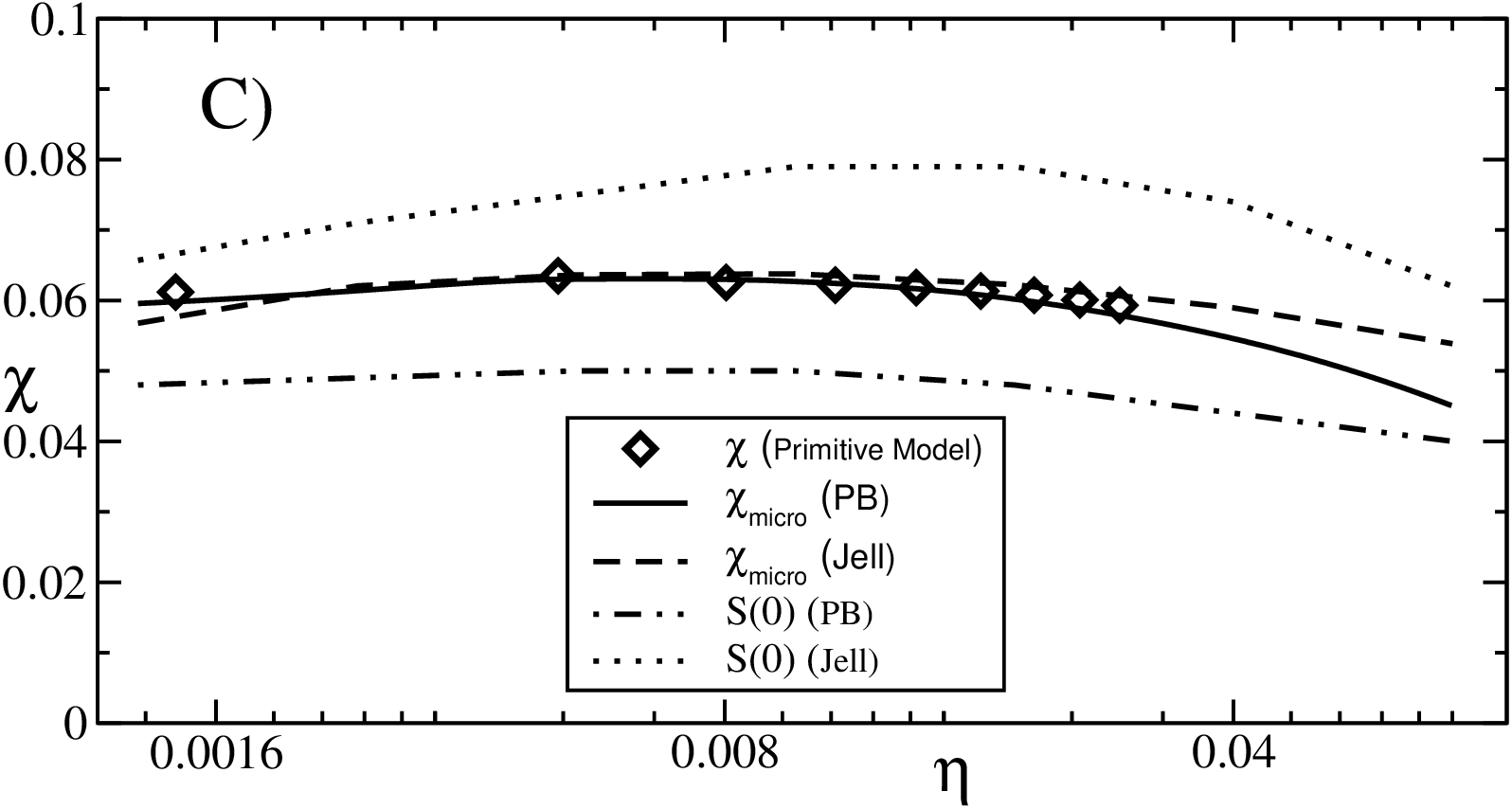}
\caption{\label{fig6} A) Osmotic pressure of a salt-free colloidal
suspension according to the primitive model simulations of Ref.
\cite{linse00}, compared to the PB-cell and Jellium predictions 
of Eq.~(\ref{eq14}). All data are computed for $Z_{\text{bare}}\,
\lambda_B/a =14.2$ and $\lambda_B/a = 0.35$. The effective charges are
not saturated. If they were, one would obtain the PB cell pressure
marked by the dashed-dotted line. B) Colloid-colloid pair-correlation
functions from Linse's primitive model simulation (symbols) and from
MC Yukawa simulations with effective PB-cell (solid lines) and Jellium
(dashed lines) parameters, see also \cite{vladimir}. From left to
right, the packing fractions are $\eta=0.08$, $0.04$, $0.01$,
$0.0025$, $0.00125$. C) Compressibilities derived from the data in A)
via Eq.~(\ref{eq4}) and from the colloid-colloid $S(0)$ as indicated.}
\end{figure}

P. Linse has carried out a full primitive model simulation of a
salt-free charge-stabilized colloidal suspension and computed the
pressure $P$ as a function of volume fraction $\eta$ \cite{linse00},
see also \cite{vladimir}. The highest colloidal charge considered in
this work does not lead to fully saturated effective charges
($Z_{\text{bare}\,}\lambda_B/a \simeq 14.2$).  Fig.~(\ref{fig6}.A)
shows the osmotic pressure as a function of $\eta$ from \cite{linse00}
and compares it to $\Pmicro$ ($=\Pi_{\text{micro}}$ if $\kappa a=0$).
$\Pmicro$ is given by Eq.~(\ref{eq14}) with either PB-cell or Jellium
model data.  For comparison, we also added the cell-model pressure one
would obtain if the charges were saturated. This figure illustrates
the quality of PB-cell and Jellium $\Pmicro$ in the de-ionized limit,
which has been repeatedly emphasized in the previous analysis.
Fig.~(\ref{fig6}.C) shows the corresponding compressibilities, and as
expected from Fig.~(\ref{fig6}.A), it can be seen that
$\chi_{\text{micro}}$ is very close to the correct primitive model
compressibility.  Figure (\ref{fig6}.C) supports our findings of the
previous sections: at low density, the Jellium effective potential
leads to a $S(0)$ that fares slightly better than its cell
counterpart.  The OCM cell and Jellium $S(0)$ respectively provide
lower and upper bounds for the true compressibility, and
correspondingly, respectively upper and lower bounds for the pair
distribution function $g(r)$ (see Fig.~(\ref{fig6}.B)).  This can be
traced back to the force distribution seen in Fig.  \ref{fig7}B) where
the Jellium effective potential leads to a slight underestimation of
the ``exact'' force distribution, while the opposite holds for cell
data. Somehow, considering an appropriate mean of Jellium and cell
effective parameters would improve the quality of the predictions.  A
step forward in this direction has recently been made in
\cite{ramon}.

%%%%%%%%%%%%%%%%%%%%%%%%%%%%%%%%%%%%%%
\section{Summary and conclusion}

The forces between colloids in a charge-stabilized colloidal
suspension are commonly approximated by a sum of pair forces
derived from the gradient of a screened Coulomb potential
$u_{\text{eff}}$ with effective charge and screening length. A number
of models can be found in literature on how to determine these
effective parameters. We here computed them either from the
Poisson-Boltzmann cell model supplemented with the Alexander {\it et al} 
recipe
\cite{alexander,trizac3}, or from the renormalized Jellium model
\cite{trizaclevin,salete}.  The effective parameters are therefore not
fitted but derived from a well defined although difficult to control
procedure. It was our main motivation to assess the relevance of this
procedure, be it in the Jellium or in the cell case. To this end, we
have performed three tests.

{\it Test 1:} We have performed {\it in situ} measures of effective
forces in typical colloidal configurations, from the solution of a
multi-center Poisson-Boltzmann theory (with $N_c=4000$ colloids). We
are not aware of similar measures in the literature \cite{LowenMulti}.  
The distribution of these forces has been compared to forces obtained from 
summing Yukawa effective forces. Under high salt conditions, the pair-wise 
and many body approaches gave very similar force distributions (the test
was performed under conditions where cell and Jellium effective
parameters are very close).  In the no salt regime, the agreement is
less quantitative but still quite good given that the system is there
always strongly correlated, with strongly overlapping
double-layers. Jellium(/cell) forces slightly
underestimate(/overestimate) the ``true'' force.  We also found
situations where the Yukawa pair-force model fails dramatically,
predicting a total force that is up to five times smaller in magnitude
than the correct many-body force (bump observed in Fig.~(\ref{fig7}.B)
at low force ratios).

{\it Test 2:} We have used the Kirkwood-Buff identity to check the
consistency in the procedure leading to the effective Yukawa potential
$u_{\text{eff}}$ and the two effective charge models. On the whole,
the consistency of $u_{\text{eff}}$ is remarkable given the simplicity
of the underlying procedures. Our findings in detail have already been
summarized at the end of section~\ref{sec:KB}.  The most important
results are:

(i) In the salt-free case, we find consistency for both
effective-charge models. The micro-ionic contributions
$\chi_{\text{micro}}$ are consistent with the OCM $S(0)$. The Jellium
potential is slightly better at low densities, but performs less than
its cell counterpart at higher volume fraction. By contrast, the
compressibility derived from $\Pocm$ is not compatible with the OCM
$S(0)$ and consequently the OCM virial pressure $\Pocm$ is far from
the true pressure $P$.

(ii) With added salt, the $\Pocm$ provides a good approximation to the
total osmotic pressure if the volume-fraction $\eta$ is lower than the
threshold value $0.2 \eta^*$, ($\eta^*$ of
eq.~(\ref{eq:etastar})). This is the case essentially if
$u_{\text{eff}}$ is density independent, i.e., at low enough colloid
density and/or high enough salt content.

(iii) At $\kappa a \neq 0$ and if $\eta \gg \eta^*$, the system should
recover a salt-free-like state: the cell rather than Jellium effective
potential leads to a structure that is more consistent with the
isothermal compressibility $\chi_{\text{micro}}$.

{\it Test 3:} We compared our data to primitive model (PM) data.
PB-cell and Jellium both lead to a $\Pmicro$ and $\chi_{\text{micro}}$
that is in excellent agreement with the corresponding PM pressure
data. The OCM cell and Jellium $S(0)$ respectively provide lower and
upper bounds for the true compressibility, and correspondingly,
respectively upper and lower bounds for the pair distribution function
$g(r)$.

Thinking in more practical terms, we finally recommend:

(i) to use our empirical value $0.2\eta^*$ to find the regions where
$\Pocm$ can be taken as a good approximation of the true pressure and
where thus the OCM picture is valid both with respect to the structure
and to the thermodynamics, 

(ii) to always compute effective parameters applying both the Jellium
and the PB model and then to take the results as an upper and lower
bound. This latter remark is based not only on the results of the
third test, but also on our observation that the correct forces
(\ref{fig7}.B) have always been located {\it between} the forces
predicted by both models.

(iii) to use the PB cell model to calculate the pressure
$P_{\rm micro}(\rm PB)$ at volume fractions above the threshold
$\eta>0.2\eta^*$. This proves to be a remarkably consistent approximation
(which does not necessarily mean correct !) of the
true pressure $P$ in this parameter region.

\section*{Acknowledgments}

It is a pleasure to thank L. Belloni, Y. Levin and I. Pagonabarraga
for fruitful discussions.  R.C.P thanks PROMEP-Mexico and CONACyT
(grant 46373/A-1) for financial support.  J.D. acknowledges the
Marie-Curie fellowship MEIF-CT-2003-501789 and reintegration grant ERG
031089.  This work has been supported in part by the NSF PFC-sponsored
Center for Theoretical Biological Physics (Grants No. PHY-0216576 and
PHY-0225630).  E.T. acknowledges the French ANR for an ACI.

%%%%%%%%%%%%%%%%%%%%%%%%%%%%%%%%%%%%%%
\appendix

\section{Effective charge models}
\label{app:effchrg}
\subsection{The Jellium model}

The Jellium model \cite{Beresford,trizaclevin,salete} assumes the
effective charges of $N_c-1$ colloidal spheres to be smeared out in
space to form a homogeneous background charge $-\rho_{back}= -
Z_{back} \rho_c$. This background charge adds to the charge
distributions of co- and counterions, $\rho_{\pm} = c_s \exp(\mp e
\beta \phi)$, in the radial PB equation, to be solved about one
central colloidal particle. Deriving now $Z_{\text{eff}}$ by comparing
the solution of the PB equation with the known far-field expression
for the electrostatic potential $\phi(r)$, one can again compute a
colloidal background charge $Z_{back}$ which differs from the previous
one. The whole procedure is iterated until self-consistency is
achieved, that is, until $Z_{\text{eff}}(Z_{back}) = Z_{back}$.  This
consistency requirement was absent in the original formulation
proposed in \cite{Beresford} ; it is however an important ingredient
to account for non-linear screening effects.  $Z_{\text{eff}}$ thus
determined is used to calculate
\begin{equation}  \label{eq8}
\kappa^{2}_{\text{eff}}=4\pi\lambda_{B}\sqrt{Z_{\text{eff}}^{2}\rho_c^{2}+4c_{s}^{2}}
\end{equation}  
which may be rewritten as 
\begin{equation}  \label{eq9}
\kappa^{4}_{\text{eff}}a^4 = 
\left( 3 \frac{Z_{\text{eff}} \lambda_B}{a} \eta \right)^2 + \kappa^4a^4
\end{equation}  
where $\eta = 4 \pi \rho_c a^3/3$ is the colloid volume fraction while
$\kappa^2 = 8 \pi \lambda_B c_s$ is the squared inverse screening length in
the reservoir.

%%%%%%%%%%%%%
\subsection{The Poisson-Boltzmann cell model}

The PB cell model \cite{alexander,trizacLang} rests on the observation that the repulsively
interacting colloids arrange their positions such that each colloid
has a region around it which is void from other colloids and which
looks rather similar for different colloids. In other words, the
Wigner-Seitz cells around two arbitrarily selected colloids are
comparable in shape and volume. One now assumes that the total charge
within each cell is exactly zero, that all cells have the same shape,
and that one may approximate this shape such that it matches the
symmetry of the colloid, i.e., spherical cells around spherical
colloids.  The cell radius $R$ is chosen in consistency with the
colloid volume fraction, and the PB equation within the cell is solved
with appropriate boundary conditions at the cell edge and the colloid
surface.  Thus, through the finiteness of the cell plus the boundary
conditions, the presence of all those colloids not inside the cell are
taken into account. For a more detailed description of the cell model
approximation see \cite{deserno}.

From the numerical solution of the PB equation, one obtains the
electrostatic potential at the cell edge $\phi_R$, and can now proceed
to compute the effective screening parameter
\begin{equation}  \label{eq11a}
\kappa^{2}_{\text{eff}}\, a^2= \kappa^{2}a^2 \cosh\phi_R
\end{equation}  
if $\kappa^{2}a^2>0$ and 
\begin{equation}  \label{eq11b}
\kappa^{2}_{\text{eff}}\,a^2= \mu^2 \exp(-\phi_R)
\end{equation}  
if $\kappa^{2}a^2=0$ where $\mu^2$ appears in the PB equation
$\nabla^2 \phi = -\mu^2 e^{-\phi}$ of the salt-free case and is
determined from the electro-neutrality condition. Then, the effective
charge following Alexander and collaborator's recipe \cite{alexander}
is given by \cite{trizac2,trizac1,trizacLang}
\begin{equation}  \label{eq12}
Z_{\text{eff}} \frac{\lambda_B}{a}
= \gamma_0 f(\kappa_{\text{eff}}\,a,\eta^{-1/3})
\end{equation}  
where $\gamma_0=\tanh \phi_R$ in the salt case and $\gamma_0=1$ in the
no-salt case, and where the function $f(x,y)$ is given by
\begin{equation}  \label{eq13}
f(x,y)=\frac{1}{x}\left\{
(x^2y -1)\sinh (xy-x) + x(y-1)\cosh (xy-x)\right\}\:.
\end{equation}  
A simple approximation valid for large bare charges is given in
\cite{trizac3}.  We emphasize here that following such a route to
define $Z_{\text{eff}}$ and $\kappa_{\text{eff}}$, a ``natural''
relation such as that embodied in Eq. (\ref{eq8}) is lost, except in
certain particular limits \cite{trizacLang} (low density, or high
density, or low charge).  We also note that it has been shown recently
that the cell model effective charge is accurately reproduced by a
dynamical rule which defines the condensed microions through a bound
on their total energy \cite{Diehl}, a criterion that may also be
considered when mean-field breaks down.

%%%%%%%%%%%%%%%%%

\begin{figure}
\mbox{}\\[1cm]
\includegraphics[width=0.48 \textwidth]{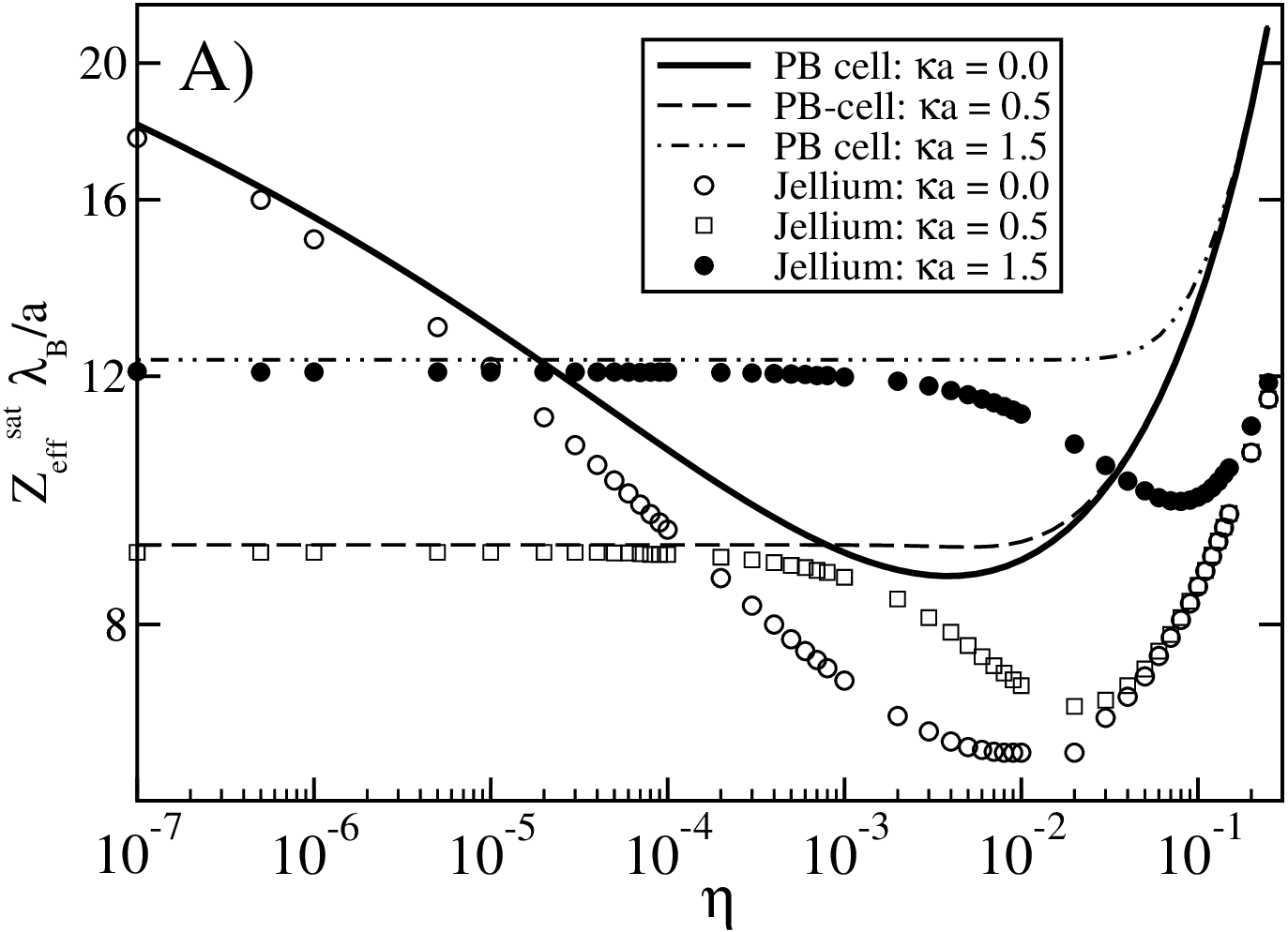}
\hfill \includegraphics[width=0.49 \textwidth]{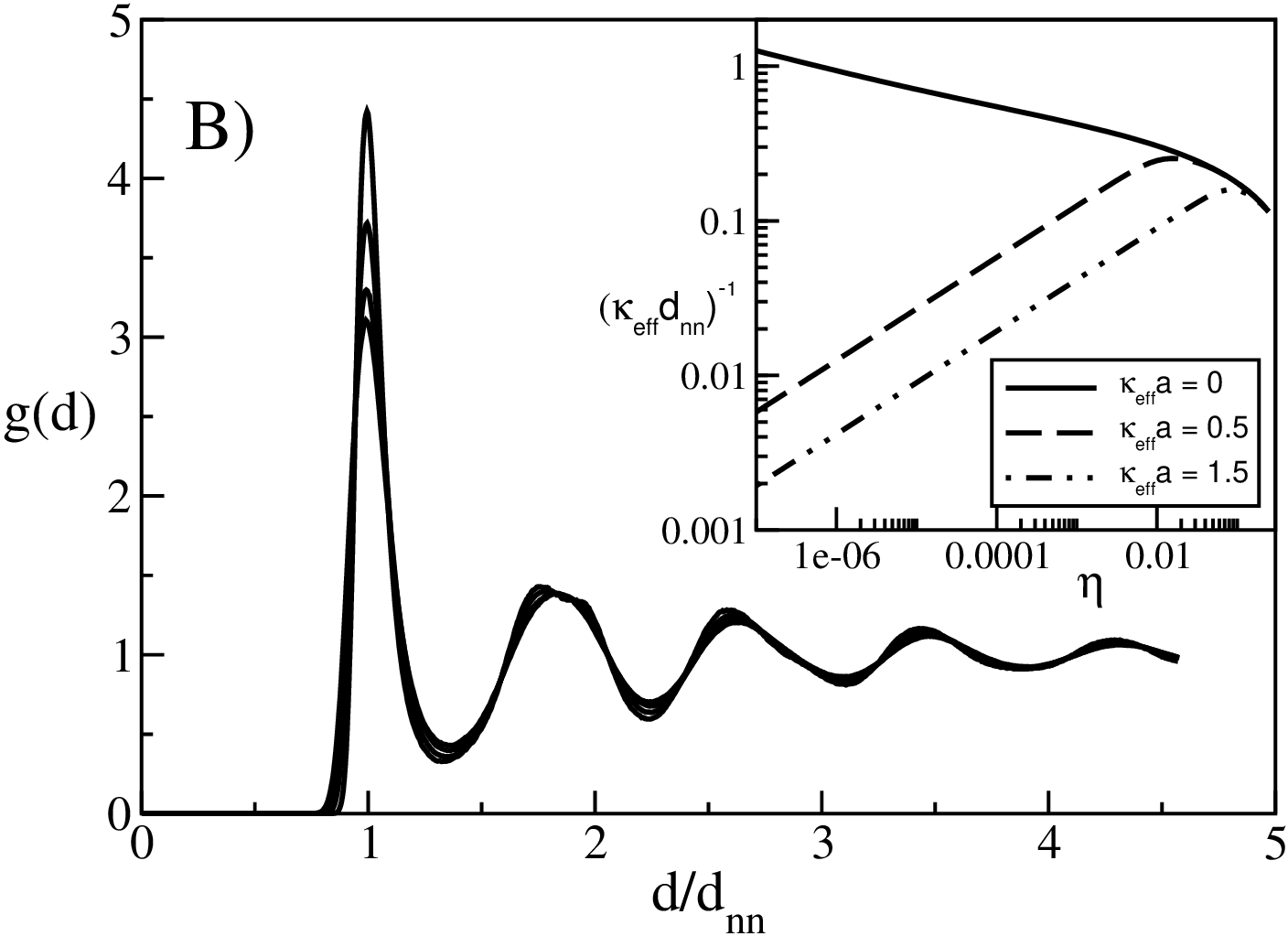}
\caption{\label{fig1} A) Effective saturated charges as a function of
the volume fraction of a charge-stabilized colloidal suspension, for
three different salinities, according to the Jellium model (symbols)
and the PB-cell model (lines). We can discriminate the salt dominated
regime $\eta<\eta^*$ and the counterion dominated regime
$\eta>\eta^*$.
%From the results presented here we can estimate
%$\eta^*_{\text PB}\approx 0.05$ and $\eta^*_{\text Jell} \approx 0.001$. 
B) The radial distribution function plotted versus the scaled
distance $d/d_{nn}$ for no-salt systems at four different volume
fractions: 0.00125 (the lowest peaks), 0.005, 0.02 and 0.08 (the
highest peaks).  There are some differences, but in the first
approximation the curves superimpose.  {\sl Inset:} The ratio
$1/(\kappa_{\text{eff}}d_{nn})$ of the double layer thickness (derived
from the cell approximation) and the mean distance between particles
as a function of the colloid volume fraction $\eta$ at three different
salt concentrations.  
%The figure serves as an demonstration of an
%already known fact: in the no salt case, upon reducing the colloid
%volume fraction, the double layer thickness increases relative to the
%mean distance between particles \cite{luc}.
}
\end{figure}

Fig.~(\ref{fig1}) serves to discuss and compare the two effective
charge models in the $(\kappa a, \eta)$-parameter space.
Fig.~(\ref{fig1}.A) compares the volume-fraction dependence of the
saturated effective charges. The change from salt-ion- to
counter-ion-dominated screening (occurring at $\eta \simeq \eta^*$)
can be recognized from the onset of a $\eta$-dependence of the
effective charges.
%: for $\kappa a =1.5$, for example, salt ions
%dominate the screening for $\eta < \eta^*_{\text PB} \approx
%0.05$ according to the cell-model predictions and for $\eta <
%\eta^*_{\text Jell} \approx 0.001$ according to the Jellium model.
For the no-salt case screening is always due to the colloidal
counter-ions, and, indeed, both models predict a strong
$\eta$-dependence of the effective charges even in the limit $\eta \to
0$.  At low $\eta$, a suspension of charged colloids with no extra
salt will always be correlated regardless of how low a volume fraction
is considered.  The reason is that the thickness of the double layer
then grows faster than the mean-distance between the particles, that
is, the ratio of the double-layer thickness $1/\kappa_{\text{eff}}$
and the mean distance $d_{nn}=\rho_c^{-1/3}$ {\em grows} with
decreasing volume fraction, as opposed to the case with external salt
where this ratio decreases (see \cite{luc} and the inset of
Fig.~\ref{fig1}.B).

Fig.~(\ref{fig1}.A) demonstrates that the two effective 
charge models agree inasmuch as the $\eta \to 0$ limiting behavior is concerned
\cite{rque10}, but disagree in the opposite limit with the Jellium
model generally predicting an earlier change from the salt-ion- to the
counterion-dominated screening and smaller effective charges. For high
$\eta$, the curves for all three values of $\kappa a$ must ultimately
converge for each model when the contribution of salt-ions to the
screening ceases to be significant. A rather special feature of the
Jellium model is the pronounced minimum in $Z_{\text{eff}}^{sat}$ at
intermediate volume fractions observed for all values of $\kappa a$,
something that is not present in the PB cell model for the salinities
investigated, but that would be observed at lower salt.

%%%%%%%%%%%%%%%%%%%%%%%
%%%%%%%%%%%%%%%%%%
\section{Numerical procedures}
\label{app:tech}

Technical details for the calculation of the effective charges and
screening parameters are given in 
\cite{trizaclevin,salete}. 
Within the OCM picture, we have performed Monte-Carlo simulations, typically 
with 10000 particles in a simulation box of side length $L_{box}$
applying periodic boundary conditions. We have carried out $5\: \cdot 10^6$
MC cycles to reach thermal equilibrium and another $2\: \cdot 10^7$ cycles
for the averages. The pressure has been obtained from Eq.~(\ref{eq3}),
the structure factor directly from
\begin{equation}  \label{eq16}
S({\bm q}) = \frac{1}{N} \left\langle
\left( \sum_i \cos({\bm q}\!\cdot\!{\bm r}_i) \right)^2
+ \left( \sum_i \sin({\bm q}\!\cdot\!{\bm r}_i)  \right)^2
\right\rangle
\end{equation}  
and $S(0)$ has then been approximated by $S(q_{min})$ where $q_{min}=2
\pi/L_{box}$ \cite{qmin}. Special care has been taken that the value converges
with the system size. As an independent check we computed the static
structure factor $S(q)$ from the $g(r)$ by Fast Fourier Transform 
and checked that the value at $q_{min}$ was the same.

Much faster than MC simulations are structure calculations using the
Ornstein-Zernike (OZ) equation \cite{hansen}, see appendix
\ref{app:integral}. This integral equation has been solved using the
well-tested Rogers-Young closure \cite{rogers}. Computing thus $g(r)$,
we obtain $S(0)$ from the integration in Eq.~(\ref{eq6}) supplemented
with finite size scaling, and the pressure from
\begin{equation}  
\label{eq18}
\beta \Pocm = \rho_c- \frac{\rho_c^2}{6} \int r g(r) \beta u'_{\text{eff}}(r) d{\bm r} \:.
\end{equation}

\begin{figure}
\includegraphics[width=0.48 \textwidth]{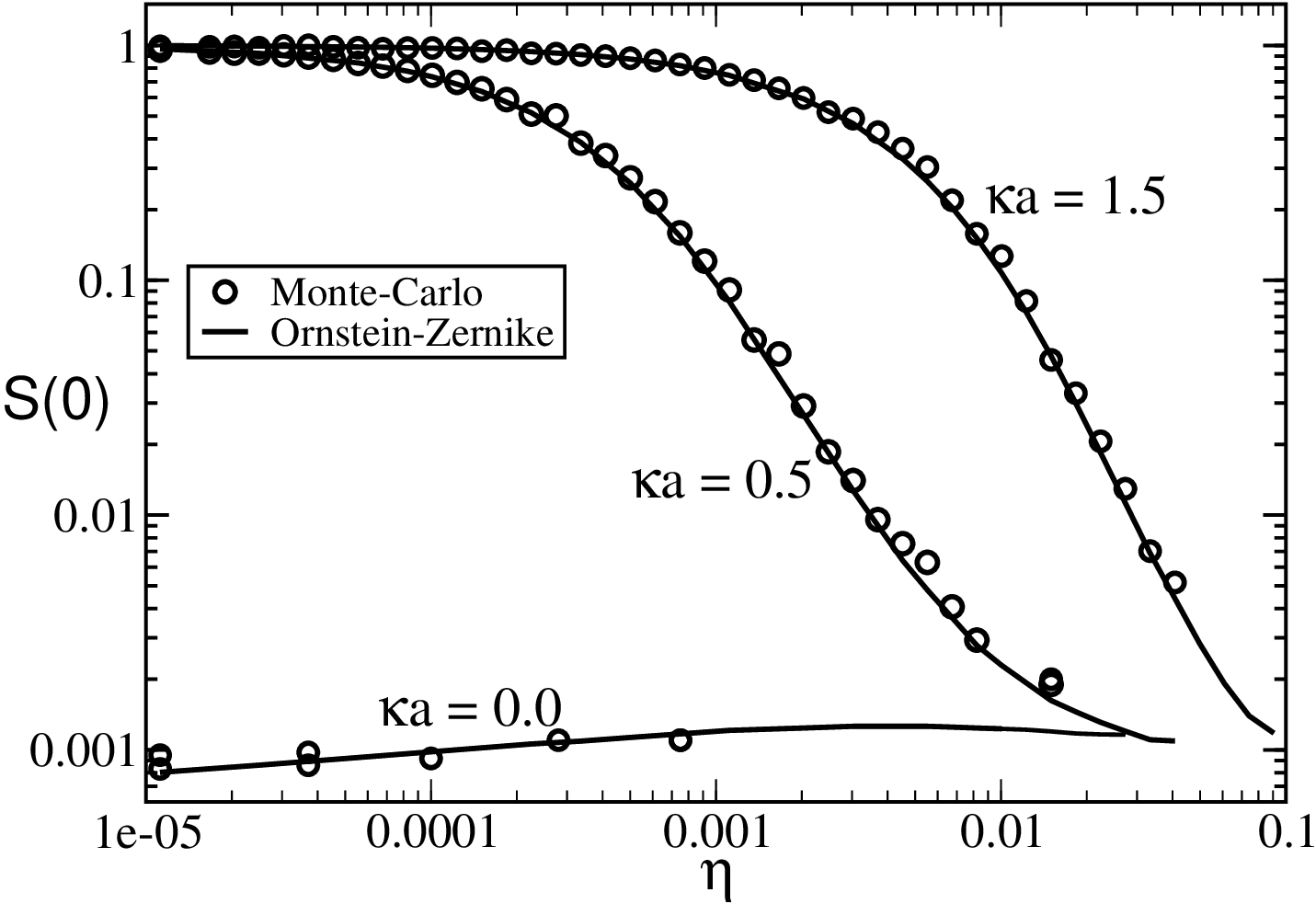}
\caption{\label{fig3}  Comparison of the large wavelength 
limit of the structure factor $S(0)$ as calculated by Monte 
Carlo simulations and by OZ integral equations method. The cell 
model approximation is used to obtain the effective parameters. 
The symbols represent the MC results, the solid lines OZ values.}
\end{figure}

To assess the validity of both Monte Carlo and integral equation
routes, we show in Fig.~(\ref{fig3}) the results of our $S(0)$
calculations in which the PB-cell effective parameters have been taken
in the Yukawa potential.  The results of the MC simulations compare
favorably with the solutions of the OZ equation for all values of
$\kappa a$ considered. Equally good agreement between the results of
both methods was found for all other curves presented in this work,
but in order not to overload the graphs, we have shown
only the OZ results.

%%%%%%%%%%%%%%%%%%%%%%%
\section{Integral equations theory}
\label{app:integral}

The structure of a fluid can be obtained by experimental techniques,
computer simulations, or by solving numerically the Ornstein-Zernike
(OZ) equation. The inhomogeneous OZ equation for a mono-component
fluid is given by \cite{hansen}
\begin{equation}
h\left(  \mathbf{r}_{1},\mathbf{r}_{2}\right)=c\left(
\mathbf{r}_{1},\mathbf{r}_{2}\right) +\int_{V}d\mathbf{r}%
_{3}c\left(  \mathbf{r}_{1},\mathbf{r}_{3}\right)  \rho\left(
\mathbf{r}_{3}\right)  h\left(  \mathbf{r}_{1},\mathbf{r}_{3}\right)
,\label{aeq1}%
\end{equation}
where $h\left( \mathbf{r}_{1},\mathbf{r}_{2}\right) $ and $c\left(
\mathbf{r}_{1},\mathbf{r}_{2}\right) $ are the total and direct
correlation functions, respectively, between a particle located at
$\mathbf{r}_{1}$ and a particle located at
$\mathbf{r}_{2}$. $\rho\left( \mathbf{r}_{3}\right) $ is the local
density of particles in the system. In an homogeneous and isotropic
system the total and direct correlation functions depend only on the
relative distance between particles and the local density takes the
average value $\rho\left( \mathbf{r}_{3}\right) =\rho$, where $\rho$
is the mean density of particles. Then, equation (\ref{aeq1}) reduces
to
\begin{equation}
h\left(  r\right)  =c\left(  r\right)  +\rho\int
_{V}d\mathbf{r}^{\prime}c\left(  
\left\vert \mathbf{r}-\mathbf{r}^{\prime}\right\vert  \right)  h\left(
\left\vert \mathbf{r}-\mathbf{r}^{\prime}\right\vert \right)  .\label{aeq2}%
\end{equation}

The total correlation function is related to the local structure of
the system by means of the relation $h\left( r\right) =g\left(
r\right) -1$, where $g\left( r\right) $ is the radial distribution
function. Also, $h\left( r\right) $ is connected with the structure
factor, $S\left( q\right) $, through the relation $S\left( q\right)
=1+\tilde{h}\left( q\right)$, where $\tilde{h}\left( q\right)$ is the
Fourier transform of $h\left( r\right) $. The structure factor can be
measured, for example, by light scattering experiments.

Eq. (\ref{aeq2}) is an integral equation with two functions,
$h\left(  r\right)  $ and $c\left(  r\right)  $, which needs an
additional relation between the total and direct correlation functions to
close the set of equations. The general closure relation for the
equation (\ref{aeq2}) is given by \cite{hansen}
\begin{equation}
h\left(  r\right)  =\exp\left[  -\beta u\left(  r\right)
+h\left(  r\right)  -c\left(  r\right)  +B\left(  r\right)
\right]  -1, \label{aeq4}%
\end{equation}
where $\beta u\left( r\right) $ is the pair potential between
particles and the function $B\left( r\right) $ is the so-called bridge
function, which depends on particle density and, in general, is
unknown. There are many approximations to the bridge function, such as
Percus-Yevick (PY), hyper-netted chain (HNC) and Rogers-Young (RY)
closure relations \cite{hansen,rogers}. It is well-known that the HNC
and RY relations work well when the interaction between particles is
only repulsive beyond the hard-core interaction. The RY approximation
\cite{rogers} is given by
\begin{equation}
h\left(  r\right)=\exp\left[  -\beta u(r)\right]  \left[
1+\frac{\exp\left[  \left(  h(r)-c(r)\right)  f(r)\right]  -1}%
{f(r)}\right]-1, \label{aeq6}%
\end{equation}
where the function $f\left( r\right) =1-\exp\left( -\alpha r\right)
$. The RY closure is a mixture between the PY and HNC closure
relations. For example, when $\alpha=0$ Eq. (\ref{aeq6}) reduces to PY
approximation. As $\alpha$ increases, $f\left( r\right)$ approaches 1,
and Eq. (\ref{aeq6}) reduces to HNC approximation. The mixture
parameter, $\alpha$, is fixed by demanding that the isothermal
compressibility from the virial route, $1-\rho\tilde{c}(q=0)=S(q=0)$,
and from the compressibility route (see Eq. (\ref{eq4})), are both
equal.  In the case of density-dependent potential of interaction, as
is the case here, it is important to compute the density derivatives
leading to the above compressibilities {\em at constant potential of
interaction}.

For a given density $\rho$ the integral equation (\ref{aeq2}),
together with the RY closure relation, is numerically solved by
converting the OZ equation in an algebraic equation. This is done by
taking the Fourier transform of Eq. (\ref{aeq2}),
\begin{equation}
\tilde{h}\left(q\right)=\frac{\tilde{c}(q)}{1-\rho \tilde{c}(q)}. \label{aeq7}%
\end{equation}

Then, by applying the inverse Fourier transform to Eq. (\ref{aeq7}) we
get the desired solution. However, for strongly interacting systems,
as in our case, the direct application of Eq. (\ref{aeq7}) gives noisy
solutions. Then, we divide the density $\rho$ in small steps of size
$\Delta \rho$. For each given sub-density we solve iteratively the OZ
by using a five-parameter version of the Ng-method \cite{ng} until the
desired density is reached. At each step of the iteration, the pair
distribution function is determined and used together with the RY
closure relation in order to compute the new direct correlation
function, $c(r)$. To ensure rapid convergence, the value of $c(r)$ at
the previous step is taken as an initial guess.

%%%%%%%%%
\section{A simple approximation for the OCM pressure}
\label{app:ocmpress}

We concentrate on the salt-free case.
With a potential given by Eq. (\ref{eq7}), a straightforward
computation of $P_{\text{ocm}}$ as given in (\ref{eq18}) leads to \cite{luc,submitted} 
\begin{eqnarray}
\label{eq:ocmvirial}
\beta P_{\text{ocm}} &=& -\frac{\rho_c^2}{6} \,\int_{r=2a}^\infty
g(r)\, \frac{d\beta u_{\text{eff}}(r)}{dr}\,r\,d^3{\bm
r}
\\
&=& \frac{2 \pi\,\rho_c^2 \,\Zeff^2 \,\lambda_B}{\kapeff^2} \left\{
1+\frac{(\kapeff a)^2}{3(1+\kapeff a)^2}\right\}
+ 
\frac{\rho_c^2 }{6} \int_{r=2a}^\infty
[g(r)-1] \,(1+\kapeff r) \, \beta  u_{\text{eff}}(r)\,d^3{\bm
r} \nonumber
\end{eqnarray} 
where the ideal gas contribution present in (\ref{eq18}) is a
small quantity (whenever $\Zeff \gg 1$ which is the case for highly 
charged colloids where $\Zeff$ is typically of order $10 a/\lambda_B$)
and has been discarded. In (\ref{eq:ocmvirial}),
the dominant term is the first one, arising from the long-range
behavior of the pair correlation function ($g\to 1$ at large
distances). In this term, the curly brackets may be safely
approximated by 1 since at low densities, $\kapeff a \ll 1$.
Therefore
\be
\beta P_{\text{ocm}} 
\,\simeq\, \frac{2 \pi\,\rho_c^2 \,\Zeff^2 \,\lambda_B}{\kapeff^2}.
\label{eq:B2}
\ee
Within the jellium, we have $\kapeff^2 = 4 \pi \lambda_B 
\Zeff \rho_c$ and $\beta \Pmicro = \Zeff \rho_c$ (that are approximately
correct within the cell model), so that
\be
\beta \Pocm \simeq \frac{1}{2} \Zeff \rho_c = \frac{1}{2} \beta \Pmicro.
\ee
Hence the factor of 2 between $\chi_{\text{ocm}}$ and
$\chi_{\text{micro}}$, roughly observed in Figs. \ref{fig4} and
\ref{fig5} in the $\kappa a=0$ case.  The fact that $\chi_{\text{ocm}}
\neq S(0)$, or in other words, that the virial and the compressibility
routes do not coincide, is a well known deficiency of
density-dependent pair potentials, see e.g. \cite{aard} for a general
discussion.  Remaining at the OCM level, improving the reliability of
the virial route -- which takes $P_{\text{ocm}}$ to approximate the
total pressure -- is possible by formally including a density
derivative of the effective potential into the forces from which the
virial is computed, see e.g. sections 4.2 of both references
\cite{luc} or \cite{aard}.  We do not follow such a route here, since
the purpose is not to test improvements of the virial route by
changing $P_{\text{ocm}}$ while neglecting $P_{\text{micro}}$, but to
test the internal consistency of a procedure (cell or jellium) that
leads both to $P_{\text{micro}}$ and to the effective potential
$u_{\text{eff}}$.

Equation (\ref{eq:B2}) also offers a clear illustration of
Kirkwood-Buff identity (\ref{eq:Pocmcomp}). To compute the right
hand-side of (\ref{eq:Pocmcomp}), we have to fix $\Zeff$ and $\kapeff$
while computing the derivative, so that \be \frac{\partial
\Pocm}{\partial \rho_c}\biggl|_{T,\text{potential}} \,\simeq\, \frac{4
\pi\,\rho_c \,\Zeff^2 \,\lambda_B}{\kapeff^2} \simeq \Zeff.  \ee On
the other hand, since $P\simeq \Pmicro$, \be \frac{\partial P
}{\partial \rho_c}\biggl|_{T,\text{salt}} \, \simeq\, \Zeff + \rho_c
\frac{\partial \Zeff}{\partial \rho_c}\biggl|_{T}, \ee where the
second term on the right hand side is negligible, as already argued in
section \ref{sec:structthermo}.  We consequently see that
Eq. (\ref{eq:Pocmcomp}) is fulfilled, within the approximations
invoked.

%%%%%%%%%%%%%%%%%%%%%%%%%%%%%%%%%%%%%


\begin{thebibliography}{10}
\bibitem{LowenMulti} A somewhat similair idea has been worked on in \cite{LowenEPL}, where a multidimensional fit has been used to find the effective parameters of the pair Yukawa interaction that optimally agree with the results of the density functional theory. However, the individual forces have not been analysed and directly compared, which is the main idea in our Test 1.
\bibitem{rque20}
We emphasize again that
the effective force felt by colloid $i$, defined as the micro-ion
average instantaneous force for a given colloidal configuration,
does not necessarily coincide with the force derived from the
gradient of the effective potential $u_{\text{eff}}$, {\em even in the hypothetical
situation where the latter is exactly known}.
\bibitem{rque10}
With salt, both models provide the same $\eta\to 0$ limit, by construction.
However, the fact that they give close results also without salt in the same
low density regime is non trivial.
\bibitem{rque4} 
To be specific, the cell or jellium model pressure $P_{\text{micro}}$
gives an excellent approximation to the full primitive model pressure
$P$ as e.g. computed in \cite{linse00}. For a coupling $z^3
Z_{\text{bare}} \lambda_B^2 / a^2 >1$, non mean-field effects become
important \cite{Netz,Levin}. Here $Z_{\text{bare}}$ is the colloidal
bare charge and $z$ the counterions' valency.
\bibitem{rque22}
In the salt-free case, the suspension behaves as if it were permanently
in the ``large'' $\eta$-limit and $S(0)$ shows a weak $\eta$
dependence only because of the $\eta$-dependence of
$Z^{sat}_{\text{eff}}$. This special feature of the zero-salt case 
can also be appreciated from the pair-correlation 
functions displayed  in Fig.(\ref{fig1}.B).  
Plotting these
functions over the distance in units of the mean separation, they largely 
resemble each other for all $\eta$ considered, indicating a
degree of particle correlation that is almost independent of $\eta$.
\bibitem{qmin}
Nicer convergence is obtained at a higher $q$ than $q_{min}$. 
The value $q=4\pi/L_{box}$ has been utilized in the 
calculations of $S(0)$. The results have converged with system size.
\end{thebibliography}
\end{document}